\documentclass[useAMS,usenatbib,usegraphicx,usedcolumn]{mn2e}

\usepackage{times} 

\defcitealias{2007MNRAS.382..109T}{T07}
\defcitealias{2011ApJ...738L..22M}{MM11}
\defcitealias{2014MNRAS.440.1634P}{PdA14}

\newcommand{\Mdyn}{\ensuremath{M_\mathrm{dyn}}}
\newcommand{\Mstar}{\ensuremath{M_\star}}
\newcommand{\Msun}{\ensuremath{\mathrm{M_\odot}}}

\newcommand{\re}{\ensuremath{r_\mathrm{e}}}
\newcommand{\rShen}{\ensuremath{r_\mathrm{Shen}(M_\star)}}
\newcommand{\kpc}{\ensuremath{\mathrm{kpc}}}

\newcommand{\sigmae}{\ensuremath{\sigma_\mathrm{e}}}
\newcommand{\kms}{\ensuremath{\mathrm{km \  s^{-1}}}}

\newcommand{\fstardyn}{\ensuremath{f_\mathrm{dyn}^\star}}

\title[Constraints on the evolution of massive galaxies]
      {Constraints on the evolutionary mechanisms of massive galaxies since
       {\boldmath $z \sim 1$} from their velocity dispersions}

\author[L. Peralta de Arriba et al.]{L.
Peralta de Arriba,$^{1,2}$\thanks{E-mail: lperalta@iac.es}
M. Balcells,$^{1,2,3}$
I. Trujillo,$^{1,2}$
J. Falc\'on-Barroso,$^{1,2}$
T. Tapia,$^{4}$
\newauthor
N. Cardiel,$^{5}$
J. Gallego,$^{5}$
R. Guzm\'an,$^{6}$
A. Hempel,$^{7}$
I. Mart\'{\i}n-Navarro,$^{1,2}$
\newauthor
P. G. P\'erez-Gonz\'alez$^{5}$
and
P. S\'anchez-Bl\'aquez$^{8}$\\
$^{1}$Instituto de Astrof\'{\i}sica de Canarias (IAC),
      E-38205 La Laguna, Tenerife, Spain\\
$^{2}$Universidad de La Laguna, Departamento de Astrof\'{\i}sica,
      E-38206 La Laguna, Tenerife, Spain\\
$^{3}$Isaac Newton Group of Telescopes,
      E-38700 Santa Cruz de La Palma, La Palma, Spain\\
$^{4}$Instituto de Astronom\'{\i}a, Universidad Nacional Aut\'onoma de M\'exico,
      Apdo. 877, Ensenada BC 22800, Mexico\\
$^{5}$Departamento de Astrof\'{\i}sica y Ciencias de la Atm\'osfera,
      Universidad Complutense de Madrid,
      E-28040 Madrid, Spain\\
$^{6}$Department of Astronomy, University of Florida,
      PO Box 112055, Gainesville, FL 32611, USA\\
$^{7}$Departamento de Ciencias F\'{\i}sicas, Universidad Andr\'es Bello,
      Av. Rep\'ublica 252, Santiago, Chile\\
$^{8}$Departamento de F\'{\i}sica Te\'orica, Universidad Aut\'onoma de Madrid,
      E-28049 Cantoblanco, Madrid, Spain}

\begin{document}

\date{Accepted 2015 July 13. Received 2015 July 9;
in original form 2015 April 1}
\setcounter{page}{704} 
\volume{453} 

\pagerange{\pageref{firstpage}--\pageref{lastpage}} \pubyear{2015} 

\maketitle

\label{firstpage}

\begin{abstract}
Several authors have reported that the dynamical masses of massive compact galaxies ($\Mstar \ga 10^{11} \  \Msun$, $\re \sim 1 \  \kpc$), computed as $\Mdyn = 5.0 \  \sigmae^2 \re / G$, are lower than their stellar masses \Mstar. In a previous study from our group, the discrepancy is interpreted as a breakdown of the assumption of homology that underlie the \Mdyn\ determinations. Here, we present new spectroscopy of six redshift $z \approx 1.0$ massive compact ellipticals from the Extended Groth Strip, obtained with the 10.4 m Gran Telescopio Canarias. We obtain velocity dispersions in the range 161--340~\kms. As found by previous studies of massive compact galaxies, our velocity dispersions are lower than the virial expectation, and all of our galaxies show $\Mdyn < \Mstar$ (assuming a Salpeter initial mass function). Adding data from the literature, we build a sample covering a range of stellar masses and compactness \emph{in a narrow redshift range $\mathit{z \approx 1.0}$}. This allows us to exclude systematic effects on the data and evolutionary effects on the galaxy population, which could have affected previous studies. We confirm that mass discrepancy scales with galaxy compactness. We use the stellar mass plane (\Mstar, \sigmae, \re) populated by our sample to constrain a generic evolution mechanism. We find that the simulations of the growth of massive ellipticals due to mergers agree with our constraints and discard the assumption of homology.
\end{abstract}


\section{Introduction} \label{sec:introduction}

The vast majority of early-type galaxies (ETGs) can safely be assumed to be in a steady stationary state. For these systems, the virial theorem holds, hence total potential and kinetic energies are balanced, such that 
\begin{equation} \label{eq:virial-theorem}
\frac{G M}{\langle r \rangle_\tau} = k_E \  \frac{\langle v^2 \rangle_\tau}{2},
\end{equation}
where $M$ is the dynamical mass, $G$ is the gravitational constant, $\langle r \rangle_\tau$ is the mean of the stellar distances to the centre of the system at time $\tau$, $\langle v^2 \rangle_\tau$ is twice the kinetic energy of the system per unit mass at time $\tau$ and $k_E = 2$ is the virialization constant.

Under the condition of structural and kinematical homology, we have $\langle r \rangle_\tau \propto \re$ and $\langle v^2 \rangle_\tau \propto \sigmae^{2}$; we then expect that ETGs verify relationships such as $M \propto \sigmae^2 \re$ and, wherever stellar mass dominates over dark matter, $\Mstar \propto \sigmae^2 \re$; here \Mstar\ is the stellar mass (computed from population synthesis modelling), \re\ is the effective (projected half-light) radius and \sigmae\ is the luminosity-weighted second moment of the line-of-sight velocity distribution (LOSVD; hereafter, velocity dispersion).

\citet{2006MNRAS.366.1126C,2013MNRAS.432.1709C} used Jeans and Schwarzschild modelling to provide a calibration of the dynamical mass for massive elliptical and lenticular galaxies in the nearby Universe, such that
\begin{equation} \label{eq:MdynTot}
\Mdyn \equiv 2 \times M_{1/2} = 5.0 \  \frac{\sigmae^{2} \re}{G}.
\end{equation}
Here $\Mdyn$ is defined as twice $M_{1/2}$, the mass enclosed in the isosurface where half the light is emitted. Therefore, the ratio \Mdyn\ to \Mstar\ provides a rigorous comparison of dynamical and stellar masses \emph{within the effective radius} (because \Mdyn\ and \Mstar\ are, respectively, twice the dynamical and stellar masses within the isosurface where half the light is emitted).

Dynamical masses of high-redshift galaxies, using equation~(\ref{eq:MdynTot}), have often led to the unphysical result $\Mdyn < \Mstar$ \citep[e.g.][]{2010ApJ...709L..58S,2011ApJ...738L..22M,2014ApJ...795..145B,2014ApJ...796...92H,2014ApJ...780..134S}. Significantly, the mass discrepancy is also found in nearby massive \emph{compact} ellipticals \citep{2012MNRAS.423..632F}.

Addressing the mass discrepancy problem, \citet[][hereinafter \citetalias{2014MNRAS.440.1634P}]{2014MNRAS.440.1634P} demonstrated that mass discrepancy does not depend on redshift, and instead scales with galaxy compactness, defined as the \re\ offset from the stellar mass--size distribution of ETGs in the nearby Universe. \citetalias{2014MNRAS.440.1634P} derived an empirical scaling of \Mdyn\ with \sigmae\ and \re\ such that
\begin{equation} \label{eq:pda2014}
\Mdyn \propto \sigmae^{3.6} \re^{0.35},
\end{equation}
which implies a breakdown of homology.

The galaxies analysed by \citetalias{2014MNRAS.440.1634P} cover a wide redshift range ($0 < z < 2.5$) and come from diverse, heterogeneous sources. In this work, we analyse the \Mdyn--\Mstar\ relationship over a narrow redshift range around $z \approx 1.0$. Hence, we exclude systematic effects on the data and evolutionary effects on the galaxy population, which could plausibly have affected previous work.

The core data set are 10 velocity dispersion measurements of 9 extremely compact massive ellipticals at $z \sim 1$, taken with the 10.4 m Gran Telescopio Canarias (GTC), 6 of which are new measurements that we describe in detail. We combine these data with velocity dispersions from the literature and draw a sample that covers stellar masses $10^{11} \ \Msun \la \Mstar < 10^{12} \ \Msun$ and effective radii ranging from `normal' to the most compact massive galaxies. This sample is ideal to show whether, at a narrow, cosmologically distant epoch, the deviation between dynamical and stellar mass follows the scaling relationship proposed by \citetalias{2014MNRAS.440.1634P}. We use a large $z \sim 0$ sample of massive galaxies to further test the predictions by comparing two snapshots in cosmic time.

The paper is organized as follows. In Section~\ref{sec:samples}, we describe the samples used in the paper: massive compact galaxies at $z\sim 1$ (Section~\ref{subsec:compact-z1}), the additional data at $z\sim 1$ (Section~\ref{subsec:other-z1}) and the nearby reference at $z\sim 0$ (Section~\ref{subsec:nyu-sample}). In Section~\ref{subsubsec:stellar-mass}, we also report the recent \Mstar\ values for our new six massive compact galaxies, extracted from the Rainbow Cosmological Surveys data base \citep[see][]{2008ApJ...675..234P,2011ApJS..193...13B,2011ApJS..193...30B}; while in Section~\ref{subsubsec:aperture-correction} we describe the aperture correction for the velocity dispersions. In Section~\ref{sec:data}, we explain the spectroscopic observations (Section~\ref{subsec:observations}), data reduction (Section~\ref{subsec:reduction}) and velocity dispersion measurements (Section~\ref{subsec:velocity-dispersions}). Accurate sky subtraction, done by implementing the method of \citet{2003PASP..115..688K} within the \textsc{reduceme} package\footnote{http://guaix.fis.ucm.es/$\sim$ncl/reduceme} \citep{1999PhDT........12C}, is demonstrated in Appendix~\ref{ap:sky-subs}, while the robustness of our velocity dispersion measurements is shown in Appendix~\ref{ap:robustness}. In Appendix~\ref{ap:spectra}, the reduced spectra and the fits are plotted. Appendix~\ref{ap:sample-z1} gathers the structural parameters for the $z \sim 1$ sample. We compare dynamical and stellar masses in Section~\ref{sec:dynamical-versus-stellar}, confirming the dependence on compactness. In Section~\ref{sec:stellar-mass-plane}, we study the relationship between \Mstar, \re\ and \sigmae\ (the stellar mass plane), and use this information to constrain a generic evolution mechanism in Section~\ref{subsec:evolution}. In Section~\ref{subsec:merger-simulations}, we compare our results with numerical simulations of dry mergers. The discussion is presented in Section~\ref{sec:discussion}, while the conclusions are given in Section~\ref{sec:conclusions}. We adopt the concordance $\Lambda$ cold dark matter cosmology ($\Omega_\mathrm{m}$=0.3, $\Omega_\Lambda$=0.7, $H_0$~=~70~km~s$^{-1}$~Mpc$^{-1}$); at $z = 1$, 1 arcsec corresponds to 8.01 \kpc. The stellar masses assume a Salpeter initial mass function (IMF); where necessary, we have used the relationships in \citet{2009MNRAS.394..774L} to convert stellar masses from the literature to this IMF.

\section{Samples} \label{sec:samples}

\begin{figure*}
  \includegraphics{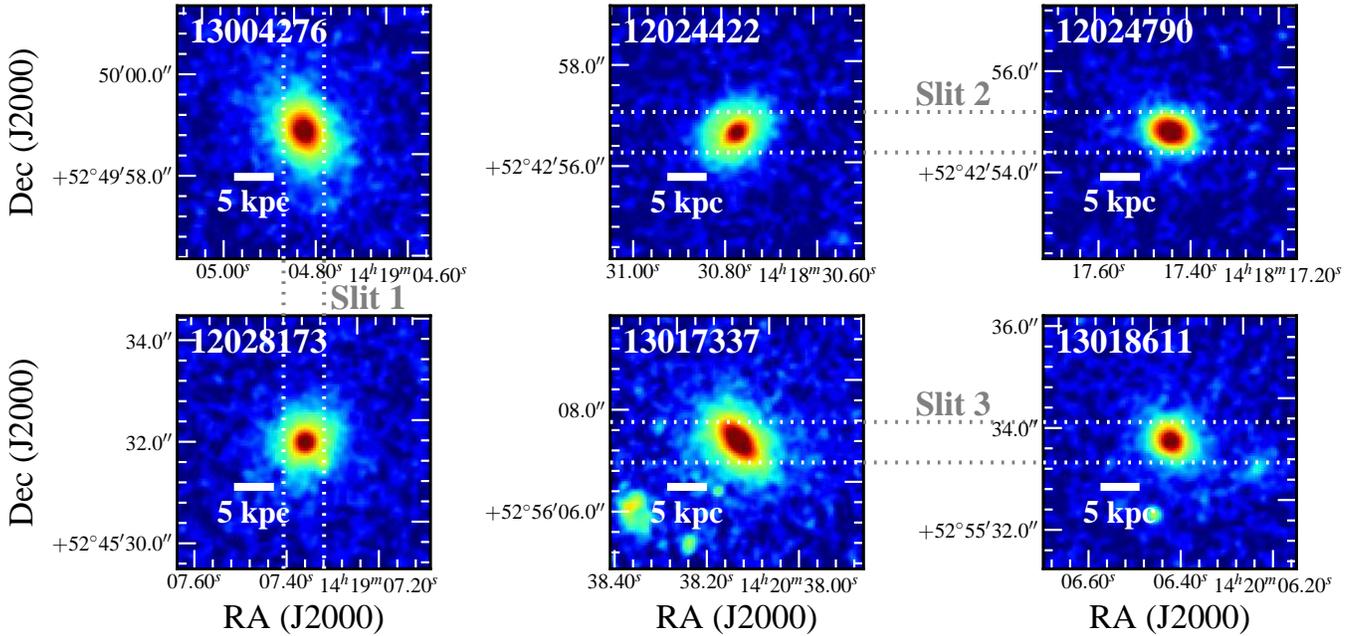}
  \caption{Images of the six massive compact galaxies at $z \sim 1$, from the Hubble Legacy Archive. These images were taken with ACS through the F814W filter ($I$ band). They are combined images and were smoothed using a Gaussian filter with a standard deviation of 1 pixel. Each panel shows a galaxy at its centre and covers 5 arcsec on a side. Galaxy identifications are shown in the top left corner of each panel. A scale bar of 5 kpc is shown in each panel. Dotted lines represent the slits used in the spectroscopic observations.}
  \label{fig:images}
\end{figure*}

The core sample for this study is a set of nine massive compact galaxies at $z \sim 1$ for which our group obtained spectroscopy on two runs with the Optical System for Imaging and low-Intermediate-Resolution Integrated Spectroscopy\footnote{http://www.gtc.iac.es/instruments/osiris} on the 10.4~m GTC. Data for four galaxies obtained in the first run are described in \citet[][hereinafter \citetalias{2011ApJ...738L..22M}]{2011ApJ...738L..22M}. The present paper fully describes the data from the second run, including sample selection and characterization (Section~\ref{subsec:compact-z1}), spectroscopy and velocity dispersion determination (Section~\ref{sec:data}).

We expand this massive compact galaxy sample with data from the literature to create a $z\sim 1$ sample with a broad coverage of the \Mstar--\re\ plane (Section~\ref{subsec:other-z1}).

We also devise a reference $z\sim 0$ sample (Section~\ref{subsec:nyu-sample}). Galaxies from both samples are similarly massive ($\Mstar \ga 10^{11} \  \Msun$) and have spheroid-like surface brightness profiles (S\'ersic index $n > 2.5$).

\subsection{Massive compact galaxies at {\boldmath $z \sim 1$}} \label{subsec:compact-z1}

Target selection for both runs was similar. We selected pairs of galaxies from the catalogue of spheroid-like ($n > 2.5$), massive ($\Mstar \ga 10^{11} \  \Msun$) galaxies in the Extended Groth Strip \citep[EGS; ][]{2007ApJ...660L...1D}, built by \citet[][hereinafter \citetalias{2007MNRAS.382..109T}]{2007MNRAS.382..109T}. This catalogue covers a redshift range from 0.2 to 2.0 and is complete in stellar mass down to $10^{11} \  \Msun$ (the details of its stellar mass estimations are presented in \citealt{2006ApJ...651..120B} and \citealt{2007MNRAS.381..962C}). Galaxies were selected in pairs given that we would use the long-slit mode of OSIRIS, and would follow the strategy of placing two targets in each slit.

The selection criteria were:
\begin{enumerate}
\item To have a S\'ersic index $n>4.0$.
\item To be closer than 4 arcmin on the sky, to allow simultaneous observation with the long slit of OSIRIS.
\item To have a spectroscopic redshift $z \approx 1.0$.
\item To be near the compact end of the size distribution at each stellar mass.
\item To have apparent magnitude $I_\mathrm{AB} < 22.2$.
\end{enumerate}

Out of the sample, four galaxies were observed in the first run and they are described in \citetalias{2011ApJ...738L..22M}. In the second run, we observed six galaxies. We chose to have one galaxy (ID 12024790) in common with the sample of \citetalias{2011ApJ...738L..22M}, to allow comparison of the velocity dispersions.

\begin{table*}
    \centering
    \begin{minipage}{154.0mm}
        \caption{Observed properties of our sample of six massive compact galaxies at redshift $\sim$1.
        (1) Galaxy identifications from the DEEP-2 Galaxy Redshift Survey \citep{2003SPIE.4834..161D,2007ApJ...660L...1D,2013ApJS..208....5N}.
        (2) Right ascensions.
        (3) Declinations.
        (4) Apparent $I$-band magnitudes in the AB system from \citet[][hereinafter \citetalias{2007MNRAS.382..109T}]{2007MNRAS.382..109T}.
        (5) Apparent $K_\mathrm{s}$-band magnitudes in the AB system from \citetalias{2007MNRAS.382..109T}.
        (6) Effective (half-light) radii along the semimajor axis from \citetalias{2007MNRAS.382..109T}.
        (7) Ellipticities from \citetalias{2007MNRAS.382..109T}.
        (8) Spectroscopic redshifts from the DEEP-2 Galaxy Redshift Survey (their absolute errors are $\la$$10^{-5}$).
        (9) Slit identifications of the spectroscopic observations.
        (10) Exposure times on target of the observations.
        (11) Rest-frame signal-to-noise ratios per angstrom in the spectra in the region of Ca\thinspace\textsc{ii} H and K (i.e. 3900--4000~\AA).}
        \label{tab:observed-obs-prop}
        \begin{tabular}{@{}lcccccccccc@{}}
        \hline
        ID       & RA                                    & Dec.                        & $I$      & $K_\mathrm{s}$ & $a_\mathrm{e}$ & $\epsilon$ & $z$     & Slit & $t_\mathrm{exp}$ & $(S/N)_\mathrm{rest}$\\
                 & (J2000)                               & (J2000)                     & (AB mag) & (AB mag)       & (arcsec)       &            &         &      & (h)              & ($\mathrm{\AA^{-1}}$)\\
        (1)      & (2)                                   & (3)                         & (4)      & (5)            & (6)            & (7)        & (8)     & (9)  & (10)             & (11)\\
        \hline
        13004276 & 14$^\mathrm{h}$19$^\mathrm{m}$04\fs81 & +52\degr49\arcmin58\farcs64 & 21.55    & 19.21          & 0.24           & 0.42       & 0.97533 & 1    & 6                & 14\\
        12028173 & 14$^\mathrm{h}$19$^\mathrm{m}$07\fs34 & +52\degr45\arcmin31\farcs59 & 22.06    & 19.59          & 0.20           & 0.20       & 0.95764 & 1    & 6                & 13\\
        12024790 & 14$^\mathrm{h}$18$^\mathrm{m}$17\fs42 & +52\degr42\arcmin54\farcs89 & 21.89    & 20.00          & 0.09           & 0.43       & 0.96572 & 2    & 4                & 19\\
        12024422 & 14$^\mathrm{h}$18$^\mathrm{m}$30\fs79 & +52\degr42\arcmin56\farcs65 & 22.19    & 19.82          & 0.26           & 0.44       & 1.02874 & 2    & 4                & 14\\
        13017337 & 14$^\mathrm{h}$20$^\mathrm{m}$38\fs08 & +52\degr56\arcmin07\farcs27 & 21.20    & 19.08          & 0.31           & 0.69       & 0.97242 & 3    & 3                & 15\\
        13018611 & 14$^\mathrm{h}$20$^\mathrm{m}$06\fs37 & +52\degr55\arcmin33\farcs61 & 22.15    & 19.99          & 0.17           & 0.26       & 1.07939 & 3    & 3                & 12\\
        \hline
        \end{tabular}
    \end{minipage}
\end{table*}

In Fig.~\ref{fig:images}, we show \textit{Hubble Space Telescope}/Advanced Camera for Surveys (ACS) images of the six galaxies of the second run. Their observed properties, as well as parameters for their spectroscopic observations, are listed in Table~\ref{tab:observed-obs-prop}. For the remaining of this paper, our sample will refer to the galaxies of the second run.

\subsubsection{Stellar masses} \label{subsubsec:stellar-mass}

For the galaxies of the second run, we upgraded the stellar masses from the \citetalias{2007MNRAS.382..109T} values to the data available in the Rainbow Cosmological Surveys data base \citep[see][]{2008ApJ...675..234P,2011ApJS..193...13B,2011ApJS..193...30B}. We chose the masses from Rainbow Cosmological Surveys data base because they have been computed using photometry with a broader spectral range (UV--FIR), while \citetalias{2007MNRAS.382..109T} used only $BRIJK$. We used stellar masses computed with spectroscopic redshifts. These masses assume a \citet{1955ApJ...121..161S} IMF, models from P\'EGASE 2.0 \citep{1997A&A...326..950F} and a \citet{2000ApJ...533..682C} extinction law. The upgraded values for these galaxies are on average 0.14 dex lower than the \citetalias{2007MNRAS.382..109T} values. This difference can be explained due to the usage of different models in their computation \citep[see][]{2009ApJ...701.1839M}. Differences are less than 1$\sigma$ for each individual galaxy.

\begin{table*}
    \centering
    \begin{minipage}{119mm}
        \caption{Derived properties of our sample of six massive compact galaxies at redshift $\sim$1.
        (1) Galaxy identifications from the DEEP-2 Galaxy Redshift Survey.
        (2) S\'ersic indices from \citetalias{2007MNRAS.382..109T}.
        (3) Circular effective (half-light) radii computed using the apparent effective radii along the semimajor axis \citepalias[from][]{2007MNRAS.382..109T}, the ellipticities \citepalias[from][]{2007MNRAS.382..109T} and the redshifts (from the DEEP-2 Galaxy Redshift Survey); their uncertainties are 10 per cent.
        (4) Stellar masses from \citetalias{2007MNRAS.382..109T}; their uncertainties are $\sim$0.2 dex.
        (5) Stellar masses from Rainbow Cosmological Surveys data base \citep[see][]{2008ApJ...675..234P,2011ApJS..193...13B,2011ApJS..193...30B}; in particular, the masses with the tag \emph{ZSPEC} of this data base, which improve the fits using spectroscopic redshifts.
        (6) Statistical errors of the stellar masses from Rainbow Cosmological Surveys data base.
        (7) Mean velocity dispersions inside the slit aperture measured in 100 Monte Carlo realizations of each spectrum (errors are estimated with their standard deviations).
        (8) Aperture correction factor \citep[computed following the prescriptions in the appendix B of][]{2013ApJ...771...85V}.
        (9) Velocity dispersions within the effective (half-light) radius.}
        \label{tab:deriv-prop}
        \begin{tabular}{@{}lccccd{1.2}ccc@{}}
        \hline
        ID       & $n$  & \re    & $\Mstar^\mathrm{T07}$   & \Mstar              & \multicolumn{1}{c}{$\Delta \Mstar$} & $\sigma$        & $\sigma / \sigmae$ & \sigmae\\
                 &      & (\kpc) & ($10^{11} \ \Msun$)     & ($10^{11} \ \Msun$) & \multicolumn{1}{c}{(dex)}           & (\kms)          &                    & (\kms)\\
        (1)      & (2)  & (3)    & (4)                     & (5)                 & \multicolumn{1}{c}{(6)}             & (7)             & (8)                & (9)\\
        \hline
        13004276 & 5.19 & 1.46   & 3.5                     & 2.0                 & 0.09                                & $320 \pm 15$    & 0.94               & $340 \pm 16$\\
        12028173 & 5.99 & 1.42   & 1.6                     & 1.2                 & 0.06                                & $214 \pm 16$    & 0.94               & $228 \pm 17$\\
        12024790 & 4.47 & 0.54   & 1.2                     & 0.9                 & 0.03                                & $241 \pm 14$    & 0.92               & $261 \pm 15$\\
        12024422 & 4.84 & 1.57   & 2.3                     & 1.2                 & 0.02                                & $225 \pm 17$    & 0.94               & $239 \pm 18$\\
        13017337 & 4.33 & 1.37   & 3.6                     & 1.8                 & 0.5                                 & $274 \pm 23$    & 0.94               & $291 \pm 25$\\
        13018611 & 4.76 & 1.19   & 1.7                     & 3.0                 & 0.05                                & $151 \pm 15$    & 0.94               & $161 \pm 16$\\
        \hline
        \end{tabular}
    \end{minipage}
\end{table*}

\begin{figure*}
  \includegraphics{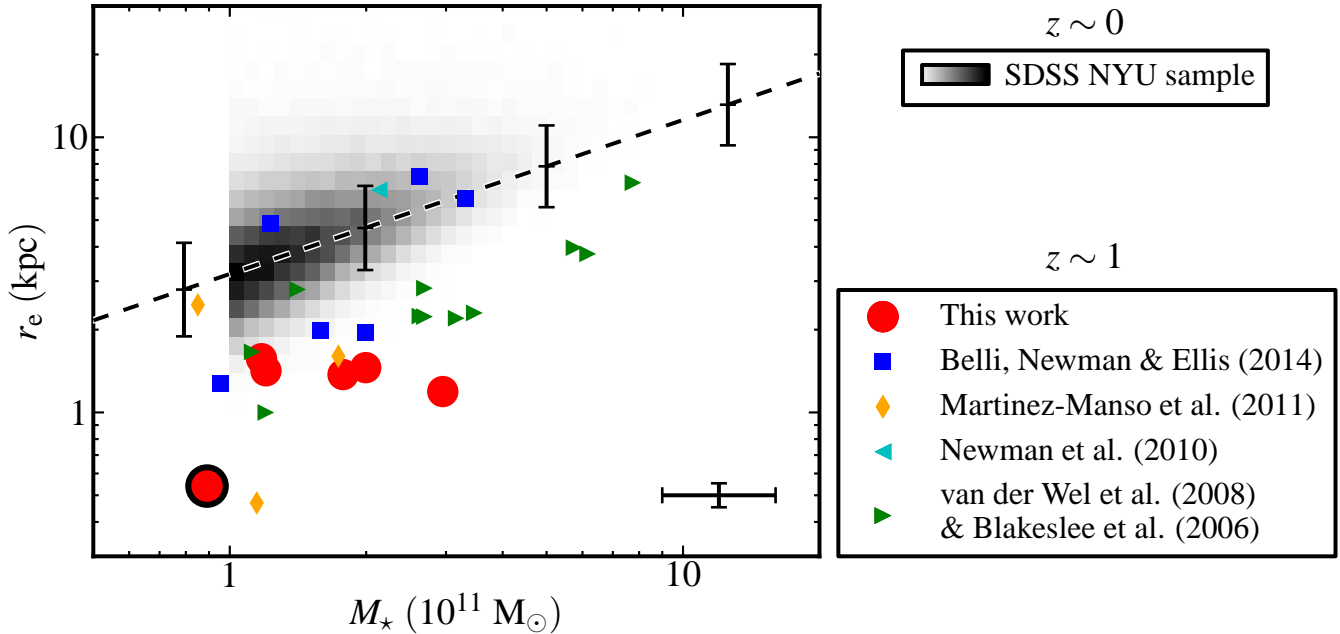}
  \caption{The stellar mass--size distribution of the spheroid-like galaxies used in this work. The sources of the $z\sim 1$ sample are given in the legend. The data point with a black edge represents the galaxy ID 12024790 also studied by \citet[][hereinafter \citetalias{2011ApJ...738L..22M}]{2011ApJ...738L..22M}. The error bar cross on the bottom right corner represents the mean errors of our six compact galaxies at $z \sim 1$. The $z\sim 0 $ sample is represented with a grey-scale giving the density of galaxies in the \Mstar--\re\ plane. The dashed line shows the relationship that spheroid-like galaxies follow in the nearby Universe \citep{2003MNRAS.343..978S}, while its error bars show the dispersion of this relationship.}
  \label{fig:stellar_mass-size}
\end{figure*}

These upgraded values for the galaxies of the second run have been included in Table~\ref{tab:deriv-prop}, which summarizes derived properties of our sample of six compact galaxies.

For the galaxies of the first run, \citetalias{2011ApJ...738L..22M} report stellar masses assuming a Chabrier IMF; we have converted their values to a Salpeter IMF. After this conversion, the stellar mass determinations in \citetalias{2011ApJ...738L..22M} and Rainbow Cosmological Surveys data base are nearly but not completely identical: \citetalias{2011ApJ...738L..22M} assume a solar metallicity, while the metallicity is not constrained in the stellar masses from the Rainbow Cosmological Surveys data base. \citetalias{2011ApJ...738L..22M} values are on average 0.11 dex lower than the values from Rainbow Cosmological Surveys data base. Nevertheless, the differences fall well within the uncertainties of each determination. We will use the stellar mass values from \citetalias{2011ApJ...738L..22M} for the galaxies of the first run in the remaining of the paper (except for the galaxy in common with the second run).

In Fig.~\ref{fig:stellar_mass-size}, we show the stellar mass--size distribution for the nine compact galaxies, together with that from the additional data described in Sections~\ref{subsec:other-z1} and \ref{subsec:nyu-sample}. Appendix~\ref{ap:sample-z1} gathers the structural parameters for the $z \sim 1$ sample, including the stellar masses from the two runs of the core sample.

\subsubsection{Aperture correction of velocity dispersions} \label{subsubsec:aperture-correction}

In this paper, we will work with velocity dispersions integrated over an aperture of one effective radius. For the galaxies of the first run, \citetalias{2011ApJ...738L..22M} report velocity dispersions corrected to an aperture diameter of 1.19 kpc. We have taken their uncorrected values (around 6 per cent lower than the 1.19 kpc aperture-corrected values) and performed a correction to one effective radius following the prescriptions in the appendix B of \citet{2013ApJ...771...85V}. These prescriptions take into account the effect of the point-spread function (PSF). To consider this fact is key due to the small apparent size of the galaxies at redshift 1. Also it is important to note that this correction can be used for apertures bigger than effective radius; while this does not occur for the aperture correction equations proposed by \citet{1995MNRAS.276.1341J} and \citet{2006MNRAS.366.1126C} (see appendix B of \citet{2013ApJ...771...85V} for a detailed explanation). This correction increases around 7 per cent their uncorrected values, i.e. the final values are similar to the 1.19 kpc aperture-corrected values. Appendix~\ref{ap:sample-z1} gathers the structural parameters for the $z \sim 1$ sample, including the velocity dispersions from the first run of the core sample.

\subsection{Additional galaxies at {\boldmath $z \sim 1$}} \label{subsec:other-z1}

We have obtained from the literature a sample of 18 massive ETGs at $z \sim 1$ ($\Mstar \ga 10^{11} \  \Msun$, $n > 2.5$ and $0.9 < z < 1.1$) with published effective radii, spectroscopic redshifts and velocity dispersions. The sources of these data are:
\begin{enumerate}
\item \citet{2014ApJ...783..117B}: six galaxies.
\item \citet{2010ApJ...717L.103N}: one galaxy \citep[extracted from the compilation made by][]{2013ApJ...771...85V}.
\item \citet{2008ApJ...688...48V} and \citet{2006ApJ...644...30B}: 11 galaxies \citep[also extracted from the compilation made by][]{2013ApJ...771...85V}.
\end{enumerate}

These publications offer the velocity dispersion values corrected to an aperture of one effective radius. They also follow the prescriptions in the appendix B of \citet{2013ApJ...771...85V}.

The stellar mass--size distribution of these $z \sim 1$ data is shown in Fig.~\ref{fig:stellar_mass-size}. All authors use our cosmology so no further corrections were needed.

\subsection{The nearby reference: massive galaxies from the SDSS NYU sample} \label{subsec:nyu-sample}

We have selected a sample of 53889 galaxies from the New York University Value-Added Galaxy Catalogue \citep{2005AJ....129.2562B}, which is based in the Sloan Digital Sky Survey \citep[hereinafter SDSS;][]{2000AJ....120.1579Y}. In what follows, we will refer this data set as SDSS NYU sample. The criteria for the selection of these galaxies were: (i) they have to be massive ($10^{11} \  \Msun < \Mstar < 10^{12} \  \Msun$), (ii) they are spheroid like ($n > 2.5$ in the $r$ band), (iii) they have redshifts around the peak of the SDSS redshift distribution ($0.05 < z < 0.11$), (iv) they have reliable velocity dispersions ($70 \  \kms < \sigma < 420 \  \kms$) and (v) their physical sizes are reliable ($0.3 \  \kpc < \re < 30 \  \kpc$ in the $r$ band).

Using equation 1 of \citet{2006MNRAS.366.1126C}, we have corrected the velocity dispersions from the fixed aperture (3 arcsec) of the SDSS fibres to the expected within one effective radius.

The stellar mass--size distribution of these data is shown in Fig.~\ref{fig:stellar_mass-size}.

\section{Data} \label{sec:data}

In this section, we explain the spectroscopic observations, data reduction and velocity dispersion measurements for the second run of six massive compact galaxies described in Section~\ref{subsec:compact-z1}.

\subsection{Spectroscopic observations} \label{subsec:observations}

Long-slit spectra for the six galaxies were obtained using the OSIRIS instrument at the 10.4 m GTC telescope. Observations were carried out in queue-scheduled service mode in 10 nights between 2012 April and August. We used the grating R2500I, which covers wavelengths from 7330 to 10000 \AA, corresponding rest-frame wavelengths at $z=1.0$ are $3665 \ \mathrm{\AA} < \lambda_\mathrm{rest} < 5000 \ \mathrm{\AA}$. The slit width was 0.8 arcsec, which delivered a resolving power of $R=2050$ at $\lambda = 8000 \ \mathrm{\AA}$ as measured from the width of the arc lines, which translates into an instrumental resolution of $\sigma_\mathrm{inst} \sim 64 \  \kms$. We binned the CCD 2$\times$2, yielding a sampling of 0.254 arcsec and 1.36 \AA\ in the spatial and spectral directions, respectively. Hence, we sampled the spectral resolution element with 2.9 pixels. Observations were carried out in dark time with seeing $\la$0.9 arcsec FWHM (full width at half-maximum).

Observations were broken in observing blocks of 1 h, each containing two 30 min exposures dithered by 7 arcsec along the slit. Total on-source integration for the three slits was 16.5 h, but 3.5 h yielded low counts, probably due to high clouds, and were unused. Exposure times on target for slits 1, 2 and 3 were 6, 4 and 3 h, respectively. Spectrophotometric standards from the Isaac Newton Group of Telescopes library\footnote{http://catserver.ing.iac.es/landscape/tn065-100/workflux.php} were observed each night.

\subsection{Data reduction} \label{subsec:reduction}

\begin{figure*}
  \includegraphics[scale=0.85]{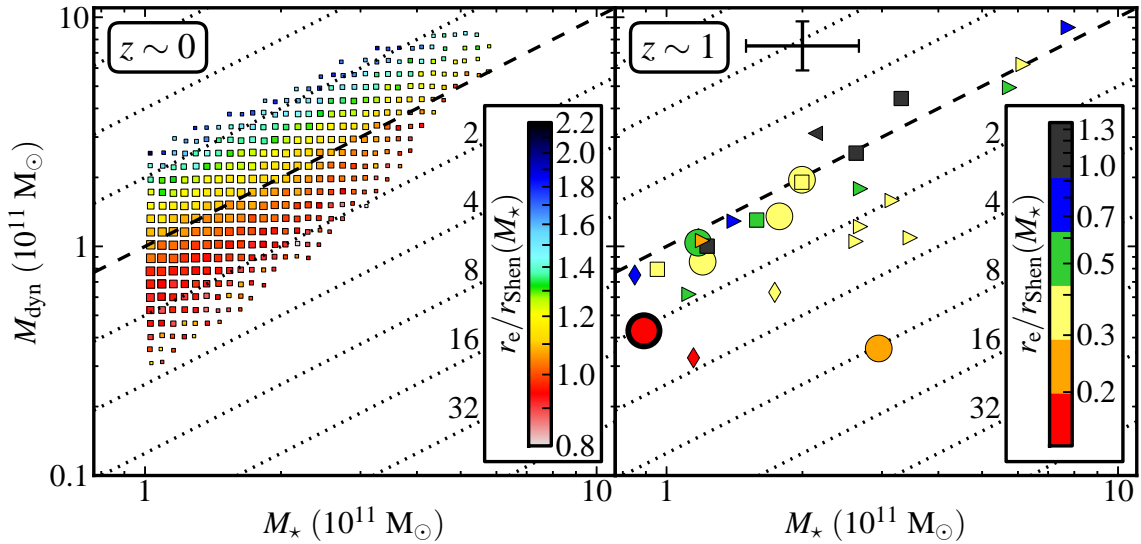}
  \caption{Stellar versus dynamical masses for the two redshift samples. Dashed line shows the identity relationship ($\Mstar = \Mdyn$). Dotted lines below/above the dashed line indicate how many times (in powers of 2) the stellar mass is greater/smaller than dynamical mass (i.e. $\Mstar = 2^{i} \Mdyn$ where $i$ is an integer). Left-hand panel shows the $z \sim 0$ sample. For this sample, the data have been binned depending on their dynamical and stellar masses. The size of each symbol in the left-hand panel scales with the number of galaxies in each bin. For clarity, bins with fewer than 10 galaxies have been omitted. In the right-hand panel, which contains galaxies with $z \sim 1$, symbol shapes as in Fig.~\ref{fig:stellar_mass-size}. In both panels, the colour of each symbol represents the compactness indicator $\re / \rShen$, defined as the \re\ offset from the stellar mass--size distribution of ETGs in the nearby Universe. The error bar cross on the top of the right-hand panel represents the mean errors of our six compact galaxies at $z \sim 1$.}
  \label{fig:mass_comparison}
\end{figure*}

We reduced our data following the standard steps (bias subtraction, flat-field correction, sky subtraction, cosmic ray removal, wavelength calibration, C-distortion correction, extinction correction, S-distortion correction, extraction and relative-flux calibration) using \textsc{reduceme} \citep{1999PhDT........12C}, a reduction package for long-slit spectra. Sky subtraction in spectra redwards of 7000 \AA\ is particularly difficult because the profiles of strong OH telluric lines, sampled by a few pixels only, get affected by spectral rectification yielding strong sky-subtraction residuals. We implemented in \textsc{reduceme} the algorithm proposed by \citet{2003PASP..115..688K}; taking advantage of the gradual shift of the wavelength solution along the slit, we build a grossly oversampled sky spectrum, which we then rebin using appropriate shifts for each CCD row. In Appendix~\ref{ap:sky-subs}, we describe the software tools developed to use this method and show examples of its effectiveness. Source extraction was carried out via straight sum within an aperture of 9 pixels (2.286 arcsec). Signal-to-noise ratios measured in the 3900--4000~\AA\ rest-frame wavelength range from 12 to 19 per angstrom; they are listed in column~11 of Table~\ref{tab:observed-obs-prop}. Relative-flux calibration and correction for telluric absorption was carried out by using the spectrophotometric standard stars.

The reduced spectra of the six galaxies are plotted in Appendix~\ref{ap:spectra}, together with the spectral fits described in Section~\ref{subsec:velocity-dispersions}. The spectra cover the rest-frame range 3800--4500 \AA. Ca\thinspace\textsc{ii} H and K, the $G$ band and several Balmer lines feature prominently, with varying relative intensities, indicating a mixture of old and intermediate-age stars. In addition, Mg\thinspace\textsc{i} 3829, 3838 \AA\ is visible in some of the spectra. The type of spectral features is similar to that found by \citetalias{2011ApJ...738L..22M}. Given the blue cut-off of our spectra, the presence of the forbidden emission line [O\thinspace\textsc{ii}] 3727~\AA\ can only be checked on two of our galaxies; no firm evidence is found on any of the two galaxies (IDs 12024422 and 13018611).

\subsection{Velocity dispersion measurements} \label{subsec:velocity-dispersions}

We computed velocity dispersions using the penalized pixel fitting (\textsc{pPXF}) method of \citet{2004PASP..116..138C}. This method characterizes the LOSVD by decomposing a spectrum as a linear superposition of templates which have been shifted considering the Doppler effect. We followed the steps described in \citet{2011MNRAS.417.1787F}, fitting the LOSVD with the first two Gauss--Hermite moments and including the computation of uncertainties using Monte Carlo realizations.

Our spectral template library comprised simple stellar populations (SSPs) from the P\'EGASE-HR library \citep{2004A&A...425..881L}. This library was selected because its high spectral resolution (0.55~\AA\ FWHM) allows us to take advantage of the quality of our spectra (which have a resolution of $\sim$2 \AA\ FWHM at rest frame). We restricted the choice of templates to SSP ages younger than age of the Universe at $z = 1$.

The use of SSPs was selected by the fact that SSPs have a more physically motivated combination of line strengths than combinations of stellar templates obtained by minimizing $\chi^2$. This is important given that Ca\thinspace\textsc{ii} H and K can bias velocity dispersion determinations because their intrinsic width varies with stellar atmospheric temperature \citep{2003ApJ...597..239G}. Because our spectra (Appendix~\ref{ap:spectra}) show prominent Balmer lines in addition to Ca\thinspace\textsc{ii} H and K, the use of SSPs prevents the fitting programme against giving a solution that depends strongly on the relative weights of the stellar template mix. We show in Appendix~\ref{ap:robustness} that the spectral fits are particularly sensitive to the choice of stellar templates.

The final fits are shown in Appendix~\ref{ap:spectra}. Although the faintest targets have higher relative sky-subtraction residuals, the main spectral features can be detected by visual inspection, and properly identified by the fitting programme.

The velocity dispersions are listed in Table~\ref{tab:deriv-prop}, together with derived properties of our sample of six massive compact galaxies. The robustness of these results is explained in Appendix~\ref{ap:robustness}. Table~\ref{tab:deriv-prop} also provides the velocity dispersions corrected to one \re\ apertures. This correction follows the prescriptions in appendix B of \citet{2013ApJ...771...85V}, which adapts the corrections proposed by \citet{1995MNRAS.276.1341J} and \citet{2006MNRAS.366.1126C} to higher apertures and takes into account the PSF effects.

\section{Dynamical versus stellar mass: discrepancy grows with compactness} \label{sec:dynamical-versus-stellar}

In this section, we compare the dynamical mass computed as $\Mdyn = K \sigmae^2 \re / G$ with $K = 5.0$ \citep{2006MNRAS.366.1126C} with the stellar masses derived using stellar population techniques.

In Fig.~\ref{fig:mass_comparison}, we plot dynamical mass versus the stellar mass for our two redshift samples. We extract two results from this figure. First, in both panels we find galaxies below the dashed line. This indicates that the unphysical result $\Mstar > \Mdyn$ occurs not only at $z \sim 1$ but also at $z\sim 0$.

Our second result relates to the dependence on galaxy compactness. The colour of each symbol in Fig.~\ref{fig:mass_comparison} has been used to indicate the compactness indicator defined as $\re / \rShen$, where $\rShen$ is the mass--size relationship for galaxies in the nearby Universe from \citet{2003MNRAS.343..978S}, written as
\begin{equation} \label{eq:shen-relation}
\rShen = r_\mathrm{Shen}^0 \left( \frac{\Mstar}{10^{11} \  \Msun} \right)^{\eta},
\end{equation}
where $r_\mathrm{Shen}^0 = 3.185 \  \kpc$ and $\eta = 0.56$. These values come from adapting the relationship in \citet{2003MNRAS.343..978S} to a Salpeter IMF following the equations from \citet{2009MNRAS.394..774L}. The compactness indicator measures the \re\ offset from the stellar mass--size distribution of ETGs in the nearby Universe.

In the left-hand panel of Fig.~\ref{fig:mass_comparison}, which shows the $z \sim 0$ galaxies from the SDSS NYU sample, we can see that red colours (low $\re / \rShen$) populate the area below the dashed line. This indicates that mass discrepancy in the nearby Universe grows with compactness. Observational errors in \re\ \citep[15 per cent;][]{2014MNRAS.444..682C} translate into a 15 per cent scatter in \Mdyn\ at a constant \Mstar, which is only one third of the observed scatter of \Mdyn\ at a constant \Mstar. This corroborates that the trend of low \Mdyn\ with compactness in Fig.~\ref{fig:mass_comparison} is real and not a result of error propagation.

At $z\sim 1$ (right-hand panel of Fig.~\ref{fig:mass_comparison}), again, the distribution of the symbol colours follows a pattern such that more compact galaxies show stronger mass discrepancies. It is worth noting that in this redshift we find higher discrepant values due to the strong size evolution of this type of galaxies (e.g. \citealt{2005ApJ...626..680D,2006MNRAS.373L..36T,2007MNRAS.374..614L,2007ApJ...671..285T}; \citetalias{2007MNRAS.382..109T}; \citealt{2007ApJ...656...66Z,2008A&A...482...21C,2008ApJ...687L..61B}).

Hence, our two samples covering narrow redshift ranges, and, therefore, largely free from redshift-dependent systematics in the mass determinations, confirm the result from \citetalias{2014MNRAS.440.1634P}, namely, that mass discrepancy grows with galaxy compactness.

We emphasize that, had we chosen a different IMF such as Chabrier, we would have alleviated the problem of galaxies showing $\Mdyn < \Mstar$, but would not have solved it. The range of IMF variations proposed by authors such as \citet{2010ApJ...709.1195T}, \citet{2012Natur.484..485C} or \citet{2013MNRAS.429L..15F} would reduce our stellar masses by at most a factor of 2. We would still have several galaxies at the low-redshift panel showing the mass anomaly. However, at high redshift, selecting other IMF will be insufficient to solve the problem. Furthermore, the authors mentioned above associate bottom-heavy IMFs with massive (high-velocity-dispersion) galaxies, which supports the choice of a Salpeter IMF made in this paper. This expectation agrees with the results from \citet{2015ApJ...798L...4M}, which found a bottom-heavy IMF for massive ETGs at $z \sim 1$.

\subsection{Weak homology: non-homology from differences in light profiles}

Several authors have proposed that a deviation from a constant value for $K$ could be originated by the different light profiles that galaxies show. Using one-component, spherical, non-rotating and isotropic models, \citet{2002A&A...386..149B} derived an equation where $K$ depends on S\'ersic index $n$ (from 6.6 to 2.2 in the interval $n \ \epsilon \ (2.5, 8)$). \citet{2006MNRAS.366.1126C} found also a similar expression assuming spherical isotropic models with S\'ersic profile (from 6.9 to 3.8 when $n$ varies from 2.5 to 8). Nevertheless, these authors found a quite constant value for their observed galaxies ($5.0 \pm 0.1$). \citet{2010ApJ...722....1T}, comparing also stellar masses from stellar population techniques with virial mass estimators, reported that following the prescriptions of \citet{2002A&A...386..149B} improved the correspondence between stellar and dynamical masses.

\begin{figure}
  \includegraphics{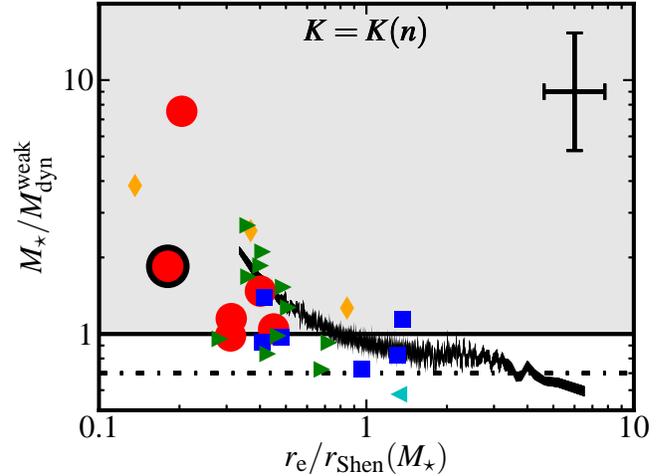}
  \caption{Correlation between $\Mstar / \Mdyn^\mathrm{weak}$ and the compactness indicator $\re / \rShen$, being $\Mdyn^\mathrm{weak}$ the dynamical mass computed with the dependence of $K$ on S\'ersic index $n$ given by equation~(\ref{eq:k_n}). Symbols for the $z \sim 1$ sample as in Fig.~\ref{fig:stellar_mass-size}. The solid black line shows the data from the $z \sim 0$ sample after grouping them in their 1000-quantiles of their $\re / \rShen$ distribution; the width of this line shows the error of the mean in each 1000-quantile. The grey area shows the unphysical region where $\Mstar > \Mdyn$. The dash--dotted line corresponds to a stellar-to-dynamical mass ratio of 0.7, derived from lensing of nearby massive ellipticals by \citet{2007ApJ...667..176G}.}
  \label{fig:weak_homology}
\end{figure}

Could this effect remove the trend between $\Mstar / \Mdyn$ and $\re / \rShen$ reported in Section~\ref{sec:dynamical-versus-stellar}? Fig.~\ref{fig:weak_homology} addresses this question showing the relation between $\Mstar / \Mdyn^\mathrm{weak}$ and $\re / \rShen$, being $\Mdyn^\mathrm{weak}$ the dynamical mass computed with a $K$ which follows the dependence on $n$ modelled by \citet{2006MNRAS.366.1126C}, i.e.
\begin{equation} \label{eq:k_n}
K(n) = 8.87 - 0.831 n + 0.0241 n^2.
\end{equation}
Thanks to this figure, we can check that the trend between $\Mstar / \Mdyn^\mathrm{weak}$ and $\re / \rShen$ remains. The trend also remains if we use the expression proposed by \citet{2002A&A...386..149B}.

\section{Stellar mass plane} \label{sec:stellar-mass-plane}

In this section, we study the relationship between \Mstar, \re\ and \sigmae. We fit a plane of the form $\Mstar \propto \sigmae^a \re^b$, with $a$ and $b$ real numbers, to the $z \sim 1$ sample. The hypotheses of virial equilibrium and homology imply a plane of the form
\begin{equation} \label{eq:virial}
\Mstar = \fstardyn \  K \  \frac{\sigmae^2 \re}{G},
\end{equation}
where
\begin{equation} \label{eq:homology}
K \textrm{ is a constant}
\end{equation}
and $\fstardyn \equiv \Mstar / \Mdyn$ may also be considered as constant, due to a random or weak dependence on \Mstar, \re\ or \sigmae. In the following, we assume $\fstardyn = 0.7$ \citep{2007ApJ...667..176G}. This assumption agrees with the results reported by \citet{2010ApJ...722....1T}, which found that \fstardyn\ depends very weakly on stellar mass at fixed S\'ersic index $n$ and varies little for the range of $n$ of our sample ($\sim$0.1 dex). As noted in Section~\ref{sec:introduction}, several works have found that equation~(\ref{eq:virial}) applies to nearby, massive ellipticals and lenticulars \citep[e.g.][]{2006MNRAS.366.1126C}. Because equation~(\ref{eq:virial}) leads to unphysical dynamical masses, \citetalias{2014MNRAS.440.1634P} proposed a modification of equation~(\ref{eq:virial}), where $K$ is allowed to vary with the compactness indicator $\re / \rShen$, such that
\begin{equation} \label{eq:k}
K = \left( \frac{\re}{\rShen} \right)^\alpha \kappa,
\end{equation}
where $\kappa$ is a constant. It is worth noting that this parametrization allows also the variation of \fstardyn with the compactness indicator $\re / \rShen$, but including it as part of the $K$ variation. These authors choose this option because they argued that the deviation from homology cannot be produced mainly by the variation of \fstardyn \citepalias[cf. fig. 9 of][]{2014MNRAS.440.1634P}.

\begin{figure*}
  \includegraphics{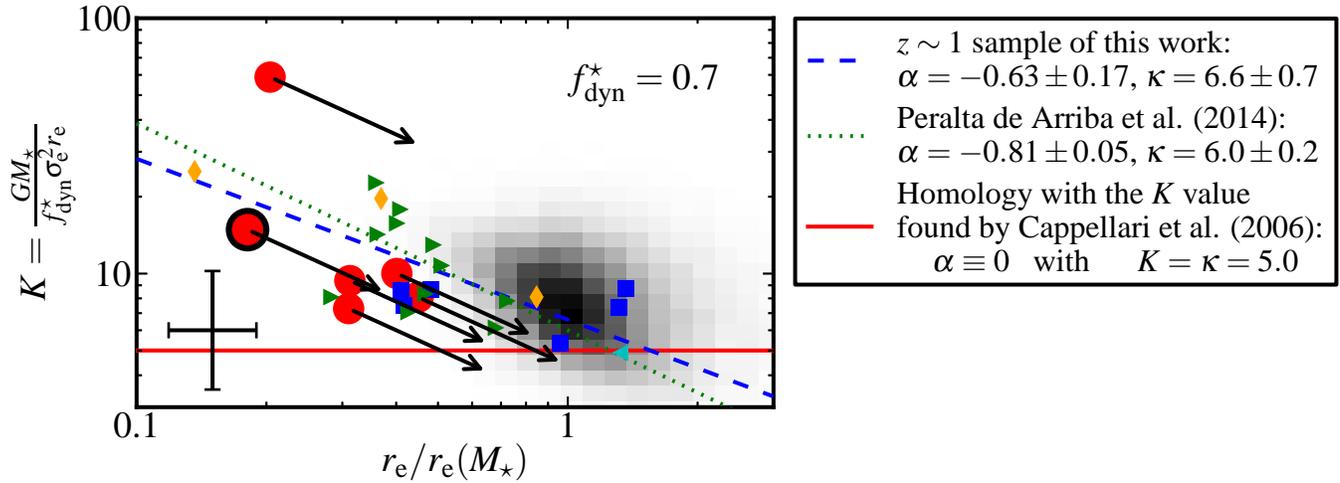}
  \caption{$K$ value versus the compactness indicator $\re / \rShen$ (i.e. the distance to nearby stellar mass--size scaling relationship). Symbols as in Fig.~\ref{fig:stellar_mass-size}. Red solid and green dotted lines show the prediction of virial theorem with homology (considering the value $K = 5.0$ found by \citealt{2006MNRAS.366.1126C}), and the prediction of \citetalias{2014MNRAS.440.1634P}. The blue dashed line shows the fit of the $z \sim 1$ sample. Black arrows show expected evolution up to $z = 0$ of the six massive compact galaxies if they would follow the equations proposed by \citet[see Section~\ref{subsec:merger-simulations}]{2015TapiaSubmitted}. The error bar cross on the bottom left corner represents the mean errors of our six compact galaxies at $z \sim 1$.}
  \label{fig:virialk-r/rshen}
\end{figure*}

Our six compact galaxies add essential data to test for variations of $K$ with compactness. In Fig.~\ref{fig:virialk-r/rshen}, we show the $K$ value versus the compactness indicator $\re / \rShen$. The red solid line represents the predictions of virial theorem with homology (equation~\ref{eq:homology}) using the value $K = 5.0$ found by \citet{2006MNRAS.366.1126C}, while we used the green dotted line to plot the predictions of \citetalias{2014MNRAS.440.1634P} (equation~\ref{eq:k}), which proposed $\alpha = -0.81 \pm 0.05$ and $\kappa = 6.0 \pm 0.2$. All of our massive compact galaxies except one follow the predictions of \citetalias{2014MNRAS.440.1634P}, while one of them shows a higher value of $K$ than both predictions. Its extreme $K$ value is caused by a low \sigmae\ ($161 \pm 16 \  \kms$); $K$ drops slightly ($K = 40.3$) but remains well above the \citetalias{2014MNRAS.440.1634P} relation, if we use the velocity dispersion from \citet[$\sigmae = 196 \pm 6 \  \kms$]{2011A&A...526A..72F}.

Fitting equation~(\ref{eq:k}) to all of our objects at $z\sim 1$, we obtain
\begin{equation} \label{eq:alpha}
\alpha = -0.63 \pm 0.17
\end{equation}
and
\begin{equation} \label{eq:kappa}
\kappa = 6.6 \pm 0.7,
\end{equation}
where the errors have been computed using the bootstrap resampling technique. This is 1$\sigma$ compatible with the results from \citetalias{2014MNRAS.440.1634P}, but it is inconsistent with the prediction of equation~(\ref{eq:homology}) ($\alpha$ is 4$\sigma$ from 0). This fit has been represented in Fig.~\ref{fig:virialk-r/rshen} with a blue dashed line. Note that in Fig.~\ref{fig:virialk-r/rshen} the regression to $z \sim 1$ (equations~\ref{eq:alpha} and \ref{eq:kappa}) matches the $z \sim 0$ sample despite the latter being excluded from the fit.

\subsection{Evolutionary constraints of stellar mass plane} \label{subsec:evolution}

In this section, we study the constraints that the stellar mass plane implies for the mechanisms which drive size evolution in this type of galaxies.

We focus on a generic mechanism in which the changes of \re\ and \sigmae\ scale as power laws of the mass growth:
\begin{equation} \label{eq:r_mass}
\frac{\re^{\mathrm{f}}}{\re^{\mathrm{i}}} =
  \left( \frac{\Mstar^{\mathrm{f}}}{\Mstar^{\mathrm{i}}} \right)^{\rho}
\end{equation}
and
\begin{equation} \label{eq:sigma_mass}
\frac{\sigmae^{\mathrm{f}}}{\sigmae^{\mathrm{i}}} =
  \left( \frac{\Mstar^{\mathrm{f}}}{\Mstar^{\mathrm{i}}} \right)^{\Sigma},
\end{equation}
where $\Mstar^{\mathrm{i}}$, $\re^{\mathrm{i}}$ and $\sigmae^{\mathrm{i}}$ denote the initial stage of an individual galaxy before the interaction of the mechanism, and the superindex `f' is analogously used to indicate the final stage. We choose such scaling relations because of their simplicity and because we expect the changes of \re\ and \sigmae\ to be strongly affected by the changes in mass.

The geometric interpretation of the above equations is that the mechanism driving size evolution moves galaxies in a well-defined direction in the three-dimensional space ($\log \Mstar$, $\log \sigmae$, $\log \re$).

If we consider that this movement has to happen inside a plane of the type defined by equations~(\ref{eq:virial}) and (\ref{eq:k}), we obtain a relationship between the exponents:
\begin{equation} \label{eq:exponents}
\alpha = \frac{\rho + 2 \Sigma - 1}{\eta - \rho},
\end{equation}
where $\eta$ is the slope of the mass--size relation for early types in the nearby Universe (equation~\ref{eq:shen-relation}). Under the virial theorem and homology (i.e. $\alpha \equiv 0$) a simpler relationship holds:
\begin{equation} \label{eq:virial_restriction}
\Sigma = \frac{1 - \rho}{2}.
\end{equation}
Additionally, relaxing the hypothesis of homology to allow a dependence of \fstardyn\ on \Mstar\ (i.e. $\fstardyn \propto \Mstar^\delta$), it leads to a shifted analogous relationship:
\begin{equation} \label{eq:relaxed_virial_restriction}
\Sigma = \frac{1 - \rho - \delta}{2}.
\end{equation}
In the following, when we consider this last case, we will assume the value $\delta \sim -0.2$ (which comes from parametrizing the variation of \fstardyn\ reported by \citet{2010ApJ...722....1T} with \Mstar).

\begin{figure*}
  \includegraphics{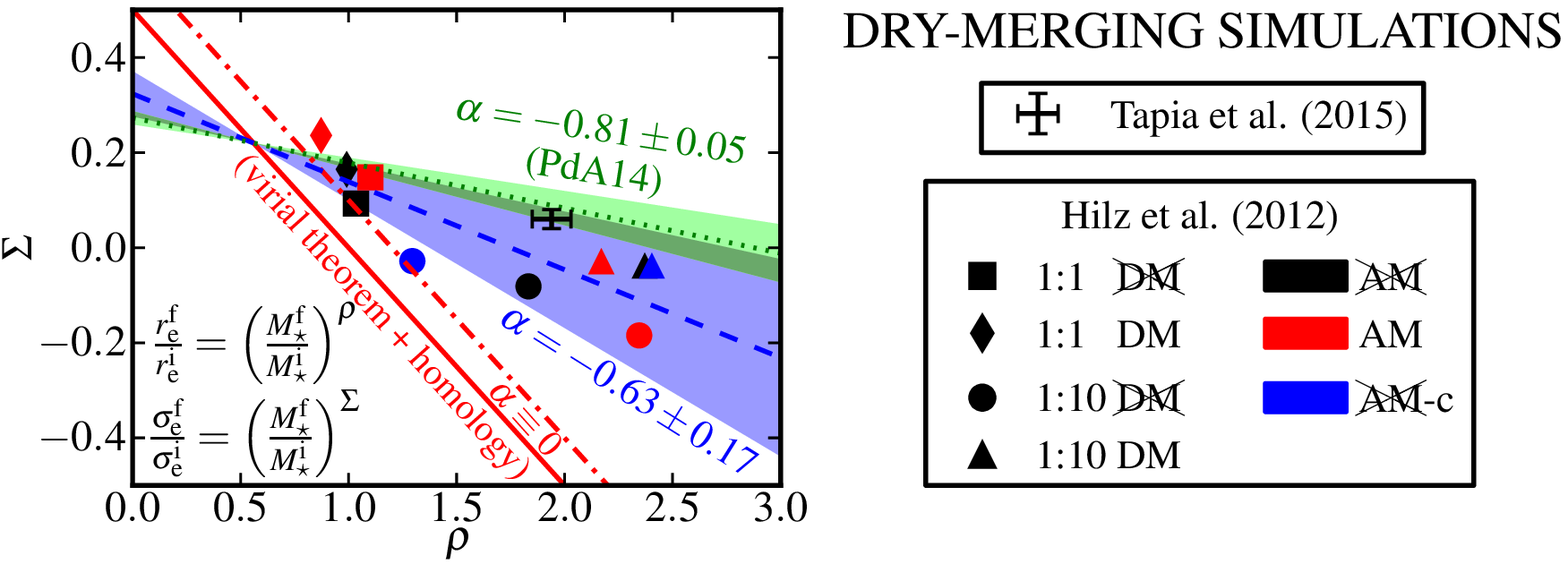}
  \caption{Stellar mass plane constraints on exponents $\rho$ and $\Sigma$ for a generic mechanism of evolution of the type $\re^{\mathrm{f}} / \re^{\mathrm{i}} = \left( \Mstar^{\mathrm{f}} / \Mstar^{\mathrm{i}} \right)^{\rho}$, $\sigmae^{\mathrm{f}} / \sigmae^{\mathrm{i}} = \left( \Mstar^{\mathrm{f}} / \Mstar^{\mathrm{i}} \right)^{\Sigma}$. Blue dashed and green dotted lines show the constraints from $\alpha$ exponents obtained from the stellar mass plane proposed in Section~\ref{sec:stellar-mass-plane} of this work and in \citetalias{2014MNRAS.440.1634P}. The shadow regions around these lines cover the 1$\sigma$ errors of these exponents. The red solid line shows the constraints from virial theorem and homology prediction. The red dash--dotted line shows the constraints from virial theorem and a \emph{relaxed} homology where stellar-to-dynamical mass ratio depends on stellar mass ($\fstardyn \propto \Mstar^{-0.2}$). Black data cross is the result found in the simulations of \citet{2015TapiaSubmitted}. The results from the simulations of \citet{2012MNRAS.425.3119H} are represented as indicated in the legend: 1:1/1:10 label corresponds to simulations of major/minor mergers, the label `DM' indicates the galaxies are embedded in matter haloes (or not if it is crossed out), the label `AM' refers to an orbit with angular momentum (or without angular momentum if it is crossed out), and the presence of the suffix `-c' means that the merging satellites were compact (i.e. satellites sizes were 50 per cent of the sizes assumed in the other simulations).}
  \label{fig:evolution_constraints-dry_mergers}
\end{figure*}

In Fig.~\ref{fig:evolution_constraints-dry_mergers}, we show the restrictions on the $\rho$--$\Sigma$ space that several stellar mass planes imply: (i) the plane predicted by virial theorem and homology (red solid line), (ii) the plane predicted by virial theorem and a \emph{relaxed} homology where $\fstardyn \propto \Mstar^{-0.2}$ (red dash--dotted line), (iii) the plane proposed by \citetalias{2014MNRAS.440.1634P} (green dotted line) and (iv) the plane obtained fitting the $z \sim 1$ sample of this work.

\subsection{Numerical simulations of dry mergers} \label{subsec:merger-simulations}

For massive compact galaxies, the cosmological evolution on the (\Mstar, \sigmae, \re) space due to their merger history can be extracted from numerical merger simulations. \citet{2015TapiaSubmitted} extracted merger histories from cosmological hydrodynamical simulations and reproduced them from $z=2.5$ to $z=0$ at high resolution, using $N$-body techniques. Initial primary galaxies had $\Mstar \sim 10^{11} \Msun$ and $\re = 1 \  \kpc$, while the merging secondaries had radii following the observational mass--size relationship at the epoch of the merger. Their simulations show that typically, a massive compact galaxy grows by a factor $\sim$2 in mass and a factor $\sim$4 in \re\ from $z=2.5$ to $z=0$.

The changes of \re\ and \sigmae\ with mass growth measured in the simulations, fitted with equations~(\ref{eq:r_mass}) and (\ref{eq:sigma_mass}), yield $\rho = 1.94 \pm 0.09$ and $\Sigma = 0.06 \pm 0.02$. These values are plotted in Fig.~\ref{fig:evolution_constraints-dry_mergers}. There is a close agreement of the results of merger simulations with the stellar mass planes of \citetalias{2014MNRAS.440.1634P} and that found in this work (equations~\ref{eq:alpha} and \ref{eq:kappa}). In contrast, the results of the merger simulations fall far from the prediction of the virial theorem and homology (even when this last assumption is relaxed to allow variations of \fstardyn\ as function of \Mstar).

\citet{2012MNRAS.425.3119H} simulated 10 series of mergers with mass ratios of 1:1 and 1:10, for a range of galaxy models (with or without dark matter haloes; varying the internal densities of the secondaries) and galaxy orbits (with or without orbital angular momentum). We have modelled the effects of these simulations with equations~(\ref{eq:r_mass}) and (\ref{eq:sigma_mass}). This allows to include the results of 10 simulations from \citet{2012MNRAS.425.3119H} in Fig.~\ref{fig:evolution_constraints-dry_mergers}. This figure shows that the action of major or minor mergers gives movements in the space (\Mstar, \sigmae, \re) which disagree with the homologous virial predictions, although major mergers and one of the minor-merger cases (mass ratio 1:10 without dark matter and with compact satellites which orbits do not have angular momentum) could be fitted with a deviation from homology where $\fstardyn \propto \Mstar^\delta$. We can check also that most of the simulations (8 out of 10) fall within the 1$\sigma$ error of the stellar mass plane proposed in this work.

The comparisons with $N$-body mergers described here show that the evolution of \re\ and \sigmae\ as \Mstar\ grows due to mergers moves massive galaxies out of the homology plane, along the mass plane proposed here. Nevertheless, a deviation from homology varying stellar-to-dynamical mass ratio with stellar mass can match the major-merger simulations and a particular case of minor mergers.

\section{Discussion} \label{sec:discussion}

It has been amply recognized that velocity dispersions contain key information to constrain the evolutionary processes linking massive compact galaxies at high redshift with nearby massive ellipticals: a simple transformation between equilibrium states by injection of dynamical energy predicts velocity dispersions around 500~\kms\ for galaxy masses $\sim10^{11} \Msun$, while galaxy growth via mergers predicts velocity dispersions to hardly evolve \citep{2010MNRAS.401.1099H}. Our measurements add to the growing body of data that shows velocity dispersions (\sigmae\ in the range 200--350 \kms) for massive compact ellipticals. To our knowledge, there remains one single case of a massive passive galaxy with reported velocity dispersion above 500 \kms, namely, a $z=2.186$ galaxy with $\Mstar = 3.2 \times 10^{11} \  \Msun$ and $\re = 0.78 \  \kpc$, for which  \citet{2009Natur.460..717V} reported a value of $510^{+165}_{-95} \  \kms$.

The comparison of our velocity dispersion measurements with those from other authors (see Appendix~\ref{ap:robustness}) indicates that uncertainties in velocity dispersions could be underestimated. \citet{2003MNRAS.339..215D} advised that the usage of Monte Carlo simulations with white noise for estimating the errors can lead to values which are almost a factor of 2 lower than realistic errors. Although we have tried to improve our errors rejecting the strong sky residuals in the fits and in the amplitude of the white noise for Monte Carlo simulations, the dependence between pixels introduced in the reduction steps should be the dominant factor for the underestimation of the errors.

The explanation of the tilt of Fundamental Plane is a classic debate of extragalactic astronomy \citep[e.g.][]{2004ApJ...600L..39T,2006ARA&A..44..141R}. Two options (or a combination of both) are possible: stellar population effects, or deviation from homology (where we include also variations of dark matter fractions). The stellar mass plane used in this work has the advantage of replacing the luminosity by the stellar mass (from stellar population techniques). This allows to address the problem in a space where the three variables are dynamically connected. Therefore, the existence of a tilt in this space implies the violation of homology.

The comparison with numerical simulations of dry mergers has worried us about some interpretation issues. For example, in the simulations from \citet{2012MNRAS.425.3119H}, although the break of homology happens in our observables (\Mstar, \sigmae, \re), it does not occur in a `theoretical' space where the variables are total bound mass, mean square speed and gravitational radius. This can be seen as an obviousness, because it only means that the equilibrium of energies predicted by virial theorem is satisfied. However, this advises that the assumption of homology is not possible for translating dry-merger simulation results from theoretical variables to observables (or even between different observables).

We have also compared our results with numerical simulations of puffing-up model. The simulations of \citet{2011MNRAS.414.3690R} predict that the three-dimensional stellar velocity dispersion and the half-mass radius are related by $\sigma_\mathrm{3D} \propto r_{1/2}^\beta$, with $\beta$ varying from $-0.49$ to $-0.83$ depending on the mass loss from gas ejection and the ejection time chosen in the simulation. As $\beta$ is not always $-1/2$, this indicates that homology can also be violated in puffing-up simulations. Our stellar mass plane predicts a law $\sigmae \propto \re^{-0.19 \pm 0.09}$ at fixed stellar mass, while the plane from \citetalias{2014MNRAS.440.1634P} gives $\sigmae \propto \re^{-0.10 \pm 0.03}$. Assuming an `ad hoc homology', i.e. $\re \propto r_{1/2}$ and $\sigmae \propto \sigma_\mathrm{3D}$, we could conclude that our stellar mass plane rejects a big contribution of puffing-up to the size evolution of ETGs. Nevertheless, when the break of homology is detected, this type of assumptions can lead to wrong results. Therefore, a puffing-up simulation which results where given in terms of observables would be useful to reject definitely the puffing-up mechanism.

Fig.~\ref{fig:evolution_constraints-dry_mergers} shows the importance of constraining observationally the stellar mass plane. The predictions from minor-merger evolution  \citep{2012MNRAS.425.3119H} fall well within the stellar mass plane presented in this paper, while they deviate from the mass plane derived in \citetalias{2014MNRAS.440.1634P}. While the set of merger models with which we compare are quite restricted, Fig.~\ref{fig:evolution_constraints-dry_mergers} shows that this type of analysis has the potential of allowing to distinguish between major and minor mergers as the dominant channel for the mass and size growth histories of today's giant galaxies. While minor mergers have often been claimed for the growth of massive compact galaxies \citep{2009MNRAS.398..898H,2012MNRAS.425.3119H,2015TapiaSubmitted}, a growth based on 1:1 to 1:5 mergers is in fact suggested by observations of the satellite distribution around massive galaxies at different redshifts \citep[e.g.][]{2014MNRAS.444..906F,2014MNRAS.442..347R}.

\section{Conclusions} \label{sec:conclusions}

The main conclusions of this paper are: 
\begin{enumerate}
\item Our measurements of velocity dispersions for six massive compact galaxies do not reveal the high values expected from a law $\Mstar \propto \sigmae^2 \re$.

\item The discrepancy between stellar and dynamical masses computed as $\Mdyn = K \sigmae^2 \re / G$ with $K = 5.0$ follows the predictions of \citetalias{2014MNRAS.440.1634P}, i.e. it scales with galaxy compactness. This implies a breakdown of homology. 

\item For our sample of massive galaxies at $z \sim 1$ which includes our six compact galaxies, we find a relationship between \Mstar, \sigmae\ and \re\ compatible with the alternative scaling law proposed by \citetalias{2014MNRAS.440.1634P}. This result is compatible with a snapshot at $z \sim 0$ with galaxies from SDSS.

\item The relationship between \Mstar, \sigmae\ and \re\ is compatible with numerical studies of the growth of massive ellipticals due to a mixture of minor and major mergers that have a cosmological framework like the one realized by \citet{2015TapiaSubmitted}. The numerical simulations of dry mergers from \citet{2012MNRAS.425.3119H} also predict a break of homology in the same direction that our stellar mass plane constrains.
\end{enumerate}

\section*{Acknowledgements}

The authors are grateful to the referee for his/her insightful and constructive review. The authors also thank J. Mart\'{\i}nez-Manso, G. Barro, A. J. Cenarro, L. Dom\'{\i}nguez-Palmero, M. Fern\'andez-Lorenzo, C. L\'opez-Sanjuan, M. Prieto, M. Cappellari, A. Cimatti, L. Ciotti, C. J. Conselice and the \emph{Traces of galaxy formation} group (http://www.iac.es/project/traces) for their collaboration during the development of this paper. LPdA was partially supported by the FPI Program by Spanish Ministry of Science and Innovation. JF-B acknowledges the support from the FP7 Marie Curie Actions of the European Commission, via the Initial Training Network DAGAL under REA grant agreement number 289313. This work has been supported by the Programa Nacional de Astronom\'{\i}a y Astrof\'{\i}sica of the Spanish Ministry of Science and Innovation under the grants AYA2009-11137, AYA2012-30717, AYA2012-31277 and AYA2013-48226-C3-1-P. Based on observations made with the Gran Telescopio Canarias (GTC), installed at the Spanish Observatorio del Roque de los Muchachos of the Instituto de Astrof\'{\i}sica de Canarias (IAC), in the island of La Palma. This work has made use of the Rainbow Cosmological Surveys data base, which is operated by the Universidad Complutense de Madrid (UCM), partnered with the University of California Observatories at Santa Cruz (UCO/Lick, UCSC). Based on observations made with the NASA/ESA Hubble Space Telescope, and obtained from the Hubble Legacy Archive, which is a collaboration between the Space Telescope Science Institute (STScI/NASA), the Space Telescope European Coordinating Facility (ST-ECF/ESA) and the Canadian Astronomy Data Centre (CADC/NRC/CSA). This research made use of \textsc{astropy}, a community-developed core \textsc{python} package for Astronomy \citep{2013A&A...558A..33A}. This research made use of \textsc{APLpy}, an open-source plotting package for \textsc{python} hosted at http://aplpy.github.com.


\appendix

\section{Sky subtraction in the data reduction} \label{ap:sky-subs}

The most critical step of our data reduction was the sky subtraction. The reason of this trouble is based on the presence of many narrow lines of sky emission on the near-infrared. In order to try to get the best possible results on this step, we developed the new software tool \textsc{pyKelsame}, which implements the sky-subtraction method proposed by \citet{2003PASP..115..688K} in the environment of \textsc{reduceme}.

In addition, in the development of \textsc{pyKelsame} we found the necessity of coding other two auxiliary tools: \textsc{reducIO} and \textsc{pySdistor}. \textsc{reducIO} is a simple module which allows to read and write files with the format of \textsc{reduceme} in a \textsc{python} environment (\textsc{reduceme} has its own format because it was designed in the 90s and then the speed of reading and writing the data played an important role). The function of \textsc{pySdistor} is to trace the spectrum of faint targets (or any) in the CCD. The algorithm of this last tool is simple: to trace the spectrum adding several channels (i.e. pixels along the spectral direction) to make easier to detect a faint object.

\begin{figure*}
  \includegraphics{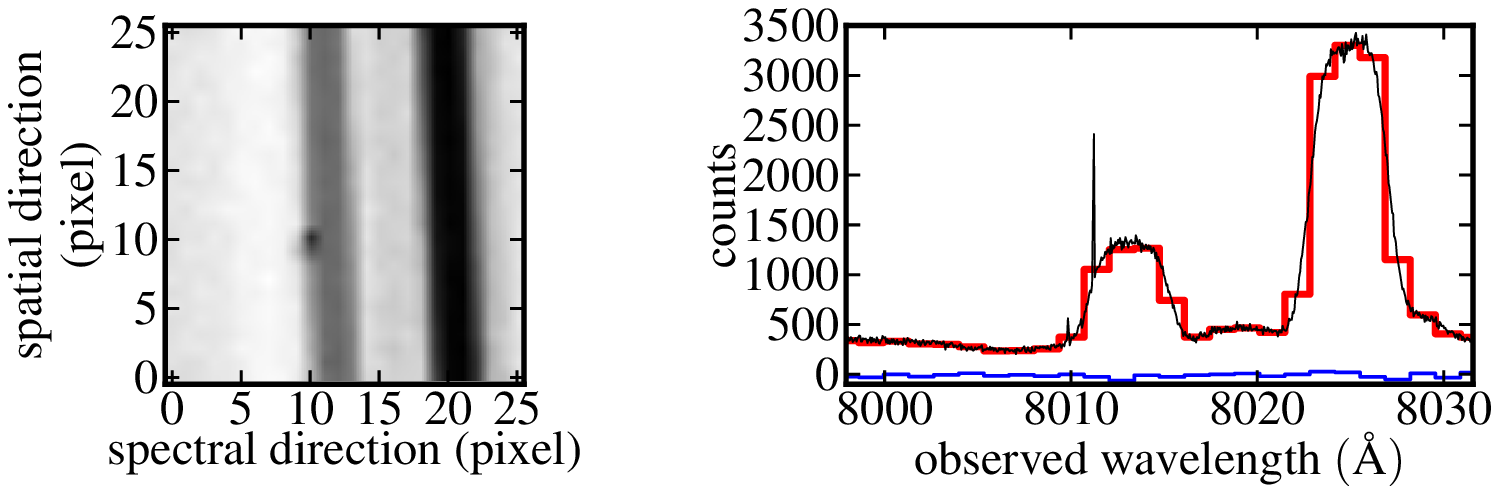}
  \caption{In the left-hand panel, we show a CCD subsection with two night-sky emission lines. In the right-hand panel, we show two spectra for these lines: one from a single CCD row (red thick line), and other using all the rows of the CCD subsection (black thin line). Note that the spectrum from all the rows of the CCD subsection takes advantage of the C-distortion for getting a spectrum more sampled than the other from single CCD row. Bottom blue medium-width line shows the residuals in the CCD row of the red thick line after the subtraction of sky model constructed from all the rows of the CCD subsection.}
  \label{fig:kelson_line}
\end{figure*}

Once we developed these two auxiliary tools, we implemented the sky-subtraction method proposed by \citet{2003PASP..115..688K}. Fig.~\ref{fig:kelson_line} illustrates the idea of this method: once we have characterized the C-distortion of the CCD (but not corrected it), we can use all the rows of the CCD to build a high-resolution spectrum of the sky. Fig.~\ref{fig:kelson_line} shows how by using every pixel of the image without rebinning (left-hand panel) we can build a high-resolution spectrum (black thin line on the right-hand panel) compared with the spectrum extracted from a single row (red thick line on the right-hand panel). It is worth noting the presence a cosmic ray in the high-resolution spectrum of this figure: it is an example of an issue which has to be solved in the construction of the sky model.

\begin{figure*}
  \includegraphics{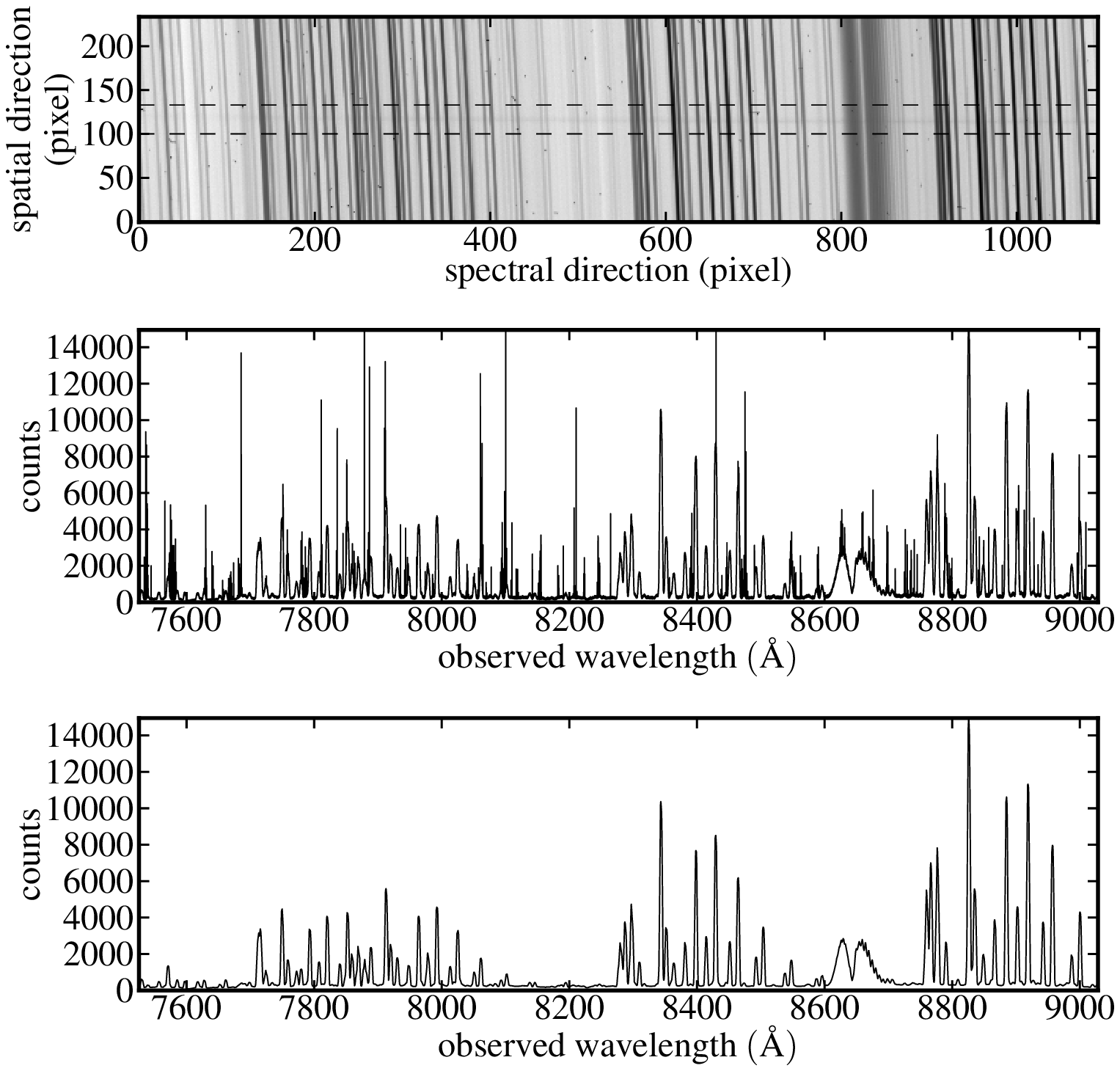}
  \caption{Top panel shows a CCD subsection; the two dashed lines indicate the region excluded for the sky modelling due to the presence of the target spectrum. Middle panel shows the spectrum build from the CCD subsection of the top panel taking advantage of the C-distortion. In bottom panel, we plot the sky model derived from the spectrum of the middle panel.}
  \label{fig:kelson_spectrum}
\end{figure*}

Fig.~\ref{fig:kelson_spectrum} illustrates the process of the construction of the sky model. In top panel, we show the first step: the CCD region to be considered in the modelling. We checked that our results improve selecting a CCD region around the target but excluding the region affected by the target spectrum (we detected the target spectrum using \textsc{pySdistor}); the dashed lines in the top panel show the excluded region. The middle panel shows the high-resolution spectrum built from the CCD region using the technique explained in the previous paragraph. It is worth noting the existence of cosmic rays in this spectrum. To reject these points, the sky model is constructed as follows: we clean the high-resolution spectrum using a median filter (or any other percentile if it was desired), and we used this clean spectrum to interpolate linearly at a desired wavelength (the clean spectrum is highly sampled, so it is not necessary a higher order interpolation). This sky model has been plotted in the bottom panel.

\begin{figure*}
  \includegraphics{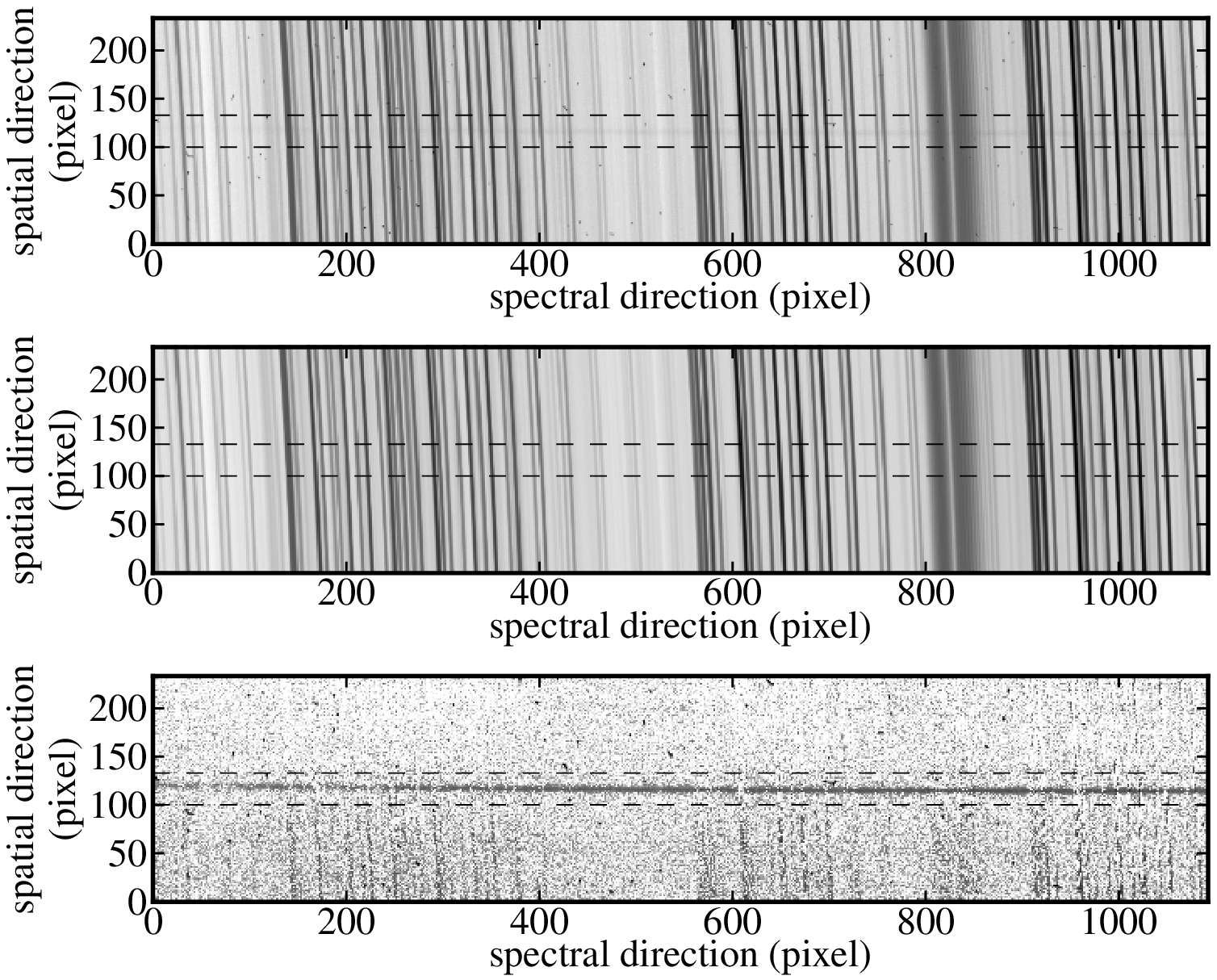}
  \caption{Top panel shows a raw CCD subsection. Middle panel shows the sky model for the CCD subsection. Bottom panel shows the CCD subsection after the sky subtraction. In each panel, dashed lines indicate the region near to the target excluded for building the sky model. Top and middle panels share the same colour bar (from 55 to 14409 counts), while another colour bar used for bottom panel (from 0.1 to 1000 counts).}
  \label{fig:kelson_ccds}
\end{figure*}

\begin{figure*}
  \includegraphics{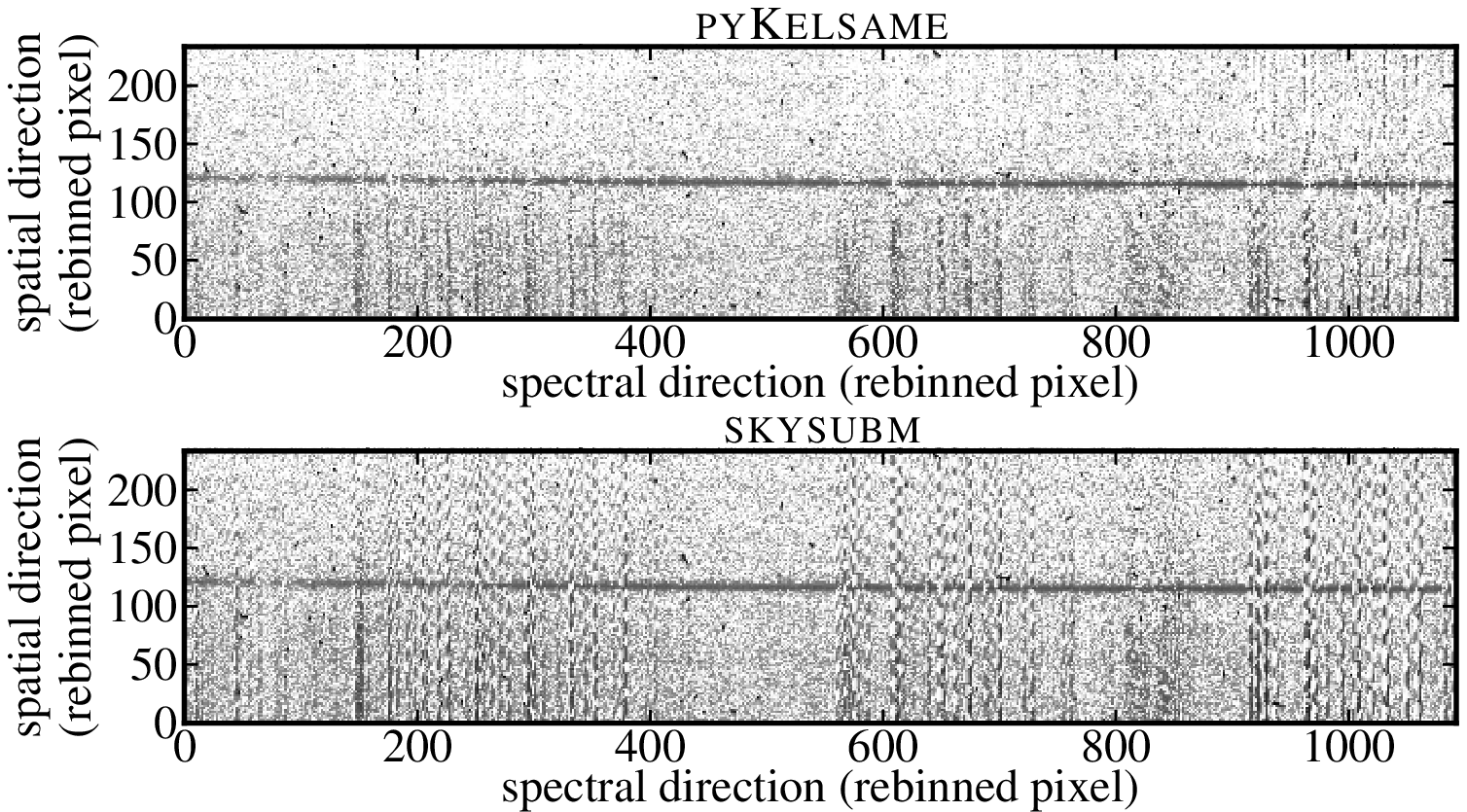}
  \caption{Top/bottom panel shows a CCD subsection where the sky subtraction was performed before/after correcting the C-distortion and using the \textsc{pyKelsame}/\textsc{skysubm} program. Both panels share the same colour bar (from 0.1 to 1000 counts).}
  \label{fig:kelson_vs_standard}
\end{figure*}

Fig.~\ref{fig:kelson_ccds} illustrates process of the sky subtraction. The top panel is the raw CCD subsection used to build the sky model. The middle panel is the evaluation of the sky model in the pixels of the raw CCD. We can check how to the cosmic rays and the target spectrum have disappeared (as it was desired). The bottom panel shows the sky-subtracted CCD. Notice that the cosmic rays are still present in the CCD, but as the sky lines have been removed, it is easier to detect them (so afterwards it will be easier to remove them).

Fig.~\ref{fig:kelson_vs_standard} compares the sky-subtraction method used in this paper with the traditional technique. Both of them have been performed in the environment of \textsc{reduceme} (using \textsc{pyKelsame} and \textsc{skysubm} programs, respectively). According to the above descriptions, the key difference between these methods is the order of subtracting sky and correcting C-distortion. For better comparison, we have also corrected the C-distortion in the panel of Fig.~\ref{fig:kelson_vs_standard} which shows the sky subtraction with \textsc{pyKelsame}: this allows that both panels display rebinned data. Checking the intensity of residuals after sky subtraction, Fig.~\ref{fig:kelson_vs_standard} confirms that the method explained in this appendix reports better results than the traditional technique.

\section{Robustness of the velocity dispersion measurements} \label{ap:robustness}

A number of tests indicate that the velocity dispersions and their errors are robust.

We obtained velocity dispersions for different combinations of SSP templates. In particular, we repeated the \textsc{pPXF} fits:
\begin{enumerate}
\item Using the SSPs from the library with ages younger than age of the Universe at $z = 1$.
\item Using the SSPs from the library with ages younger than age of the Universe at $z = 1$ and the option of regularization of \textsc{pPXF}.
\item Using the whole SSP library.
\item Using the whole SSP library, and masking H\thinspace$\gamma$ and H\thinspace$\delta$ lines.
\item Using the single SSP which alone best fits each spectrum.
\end{enumerate}
For all galaxies, the results from all the tests were in agreement with each other within their errors, being the mean standard deviation 9 \kms. In the case (iii), we tested for the possibility that residual emission might be partially filling the H\thinspace$\gamma$ and H\thinspace$\delta$ lines; because most of our spectra do not reach the [O\thinspace\textsc{ii}] 3727 \AA\ line, star formation could easily go unnoticed and affect the velocity dispersion measurements. We therefore compared our results with those from fits in which these lines were masked out.

We further tested out velocity dispersion results by carrying out the \textsc{pPXF} fits using a stellar library instead of SSPs. We used a library with 193 stars from the Indo-US Library of Coud\'e Feed Stellar Spectra \citep{2004ApJS..152..251V}. The selection of stars was performed to cover a wide range of effective temperatures, metallicities and surface gravities. This library was chosen because its high spectral resolution (1.36 \AA\ FWHM) allows us to take advantage of the quality of our spectra ($\sim$2 \AA\ FWHM at rest frame). The stellar library has the disadvantage of being less physical motivated, but, on the other hand, it provides wider spectral coverage by reaching down to 3460 \AA, as compared to the P\'EGASE-HR SSP library, which cuts at 3900 \AA. These fits were carried out masking and without masking H\thinspace$\gamma$ and H\thinspace$\delta$ lines. In all these cases, we obtain similar results: the mean of the differences is 0 and their standard deviation is 7 \kms.

The results from fitting stellar templates are very similar to those from SSPs for four of the galaxies. For the other two, the fits with stellar templates give lower, marginally inconsistent dispersions. Specifically, for the galaxy ID 13018611, we obtained $\sigma = 118 \pm 22 \ \kms$ using the stellar library and $\sigma = 151 \pm 15 \ \kms$ using SSPs (not masking the H\thinspace$\gamma$ and H\thinspace$\delta$ lines in both cases). Similarly, for the galaxy ID 12028173, using the stellar library we obtained $180 \pm 8$ and $214 \pm 16 \ \kms$ using SSPs (bootstrap errors). As discussed in Section~\ref{subsec:velocity-dispersions}, we believe the fits to stellar templates are more uncertain given the intrinsic width of the Ca\thinspace\textsc{ii} H and K lines for some stellar temperatures, and hence we choose the solutions given by SSP fits. Nevertheless, adopting the dispersions obtained with stellar templates would not change the conclusions from this work.

Finally, we have compared our results with those from other authors. We have the galaxy ID 12024790 in common with \citetalias{2011ApJ...738L..22M}. That authors reports $\sigmae = 160 \pm 10 \  \kms$, which is significantly lower than our value, $\sigmae = 261 \pm 15 \  \kms$. Applying the masking they use to exclude regions affected by telluric absorption, and cutting the redder part of our spectrum to their red cut-off we could not reproduce their result: we obtained $\sigmae = 258 \pm 22 \  \kms$. In order to explore the origin of the difference, we repeated the measurement over the \citetalias{2011ApJ...738L..22M} spectrum (albeit without relative-flux calibration) using \textsc{pPXF} and the above libraries (P\'EGASE-HR and Indo-US). Using the P\'EGASE-HR library, we performed the fit for a wide range (1--13) of degrees of the Legendre polynomial used to correct the continuum shape and found a strong dependence with this parameter obtaining values from 198 to 310 \kms. Doing the same exercise with the Indo-US library, we did not find a clear dependence with the polynomial degree and the results cover a narrower range: 227--259 \kms. We also performed these tests in our spectrum for this galaxy (using the same wavelength ranges and masking the same regions), and obtained the ranges 250--264 and 239--277 \kms\ for P\'EGASE-HR and Indo-US libraries, respectively. All these tests except one produce stable results similar to our measurement, while the unstable test corresponds to a similar situation to the published by \citetalias{2011ApJ...738L..22M}: their spectrum and SSP templates. Additional checks on our own spectra included a verification of the instrumental dispersion and its variation from night to night: measurement of the width of the telluric lines in the region of the kinematic signatures shows a very stable configuration ($3.88 \pm 0.06 \  \mathrm{\AA}$ FWHMs). We conclude that the origin of the difference with \citetalias{2011ApJ...738L..22M} is unclear, and might be related to the use of different velocity dispersion algorithms; we have used our measurement in the paper.

We share the galaxy ID 13018611 with the sample of \citet{2011A&A...526A..72F}. For this galaxy, they report $\sigmae = 196 \pm 6 \  \kms$, while we find $\sigmae = 161 \pm 16 \ \kms$. Their velocity dispersion measurement shows a dependence on the spectral range of the fit: $173 \  \kms$ for the range 3829--4129 \AA, $184 \  \kms$ for 3775--4178 \AA\ and $195 \  \kms$ for 3725--4227 \AA\ (Fern\'andez Lorenzo, private communication). The first of these three ranges is the closest to the one used by us, and, for that fitting range, the two determinations agree within the errors. In any case, the differences found show that the true uncertainties in the velocity dispersion determinations can be underestimated using the standard techniques (Monte Carlo simulations with white noise); this fact has been already proven by \citet{2003MNRAS.339..215D}. We have used our value in the paper for consistency, and emphasize that using their value would not modify the conclusions of this work.

\section{\textsc{\lowercase{p}PXF} fitting results} \label{ap:spectra}

\begin{figure*}
  \includegraphics{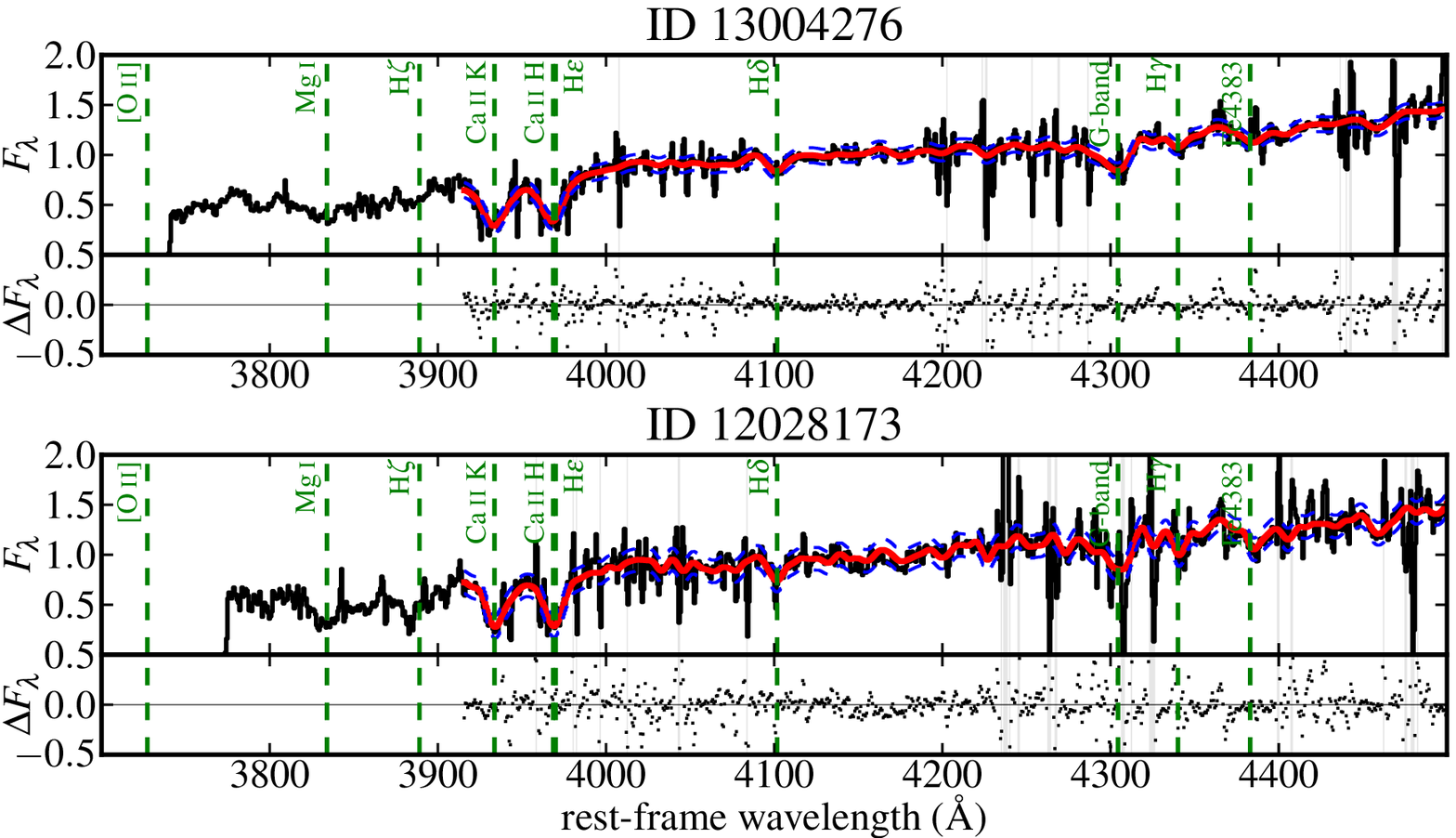}
  \caption{\textsc{pPXF} fitting results for the targets in the slit 1 of the observations. On top panels black thin solid lines show the spectra, red thick solid lines represent \textsc{pPXF} fits, and blue dashed lines are 1$\sigma$ deviations from the \textsc{pPXF} fits. On bottom panels we have plotted the residuals with black dots. Units are arbitrary. Grey regions indicate those which were masked in the \textsc{pPXF} fits. We have marked some spectral features with green dashed vertical lines.}
  \label{fig:spectra-slit1}
\end{figure*}

\begin{figure*}
  \includegraphics{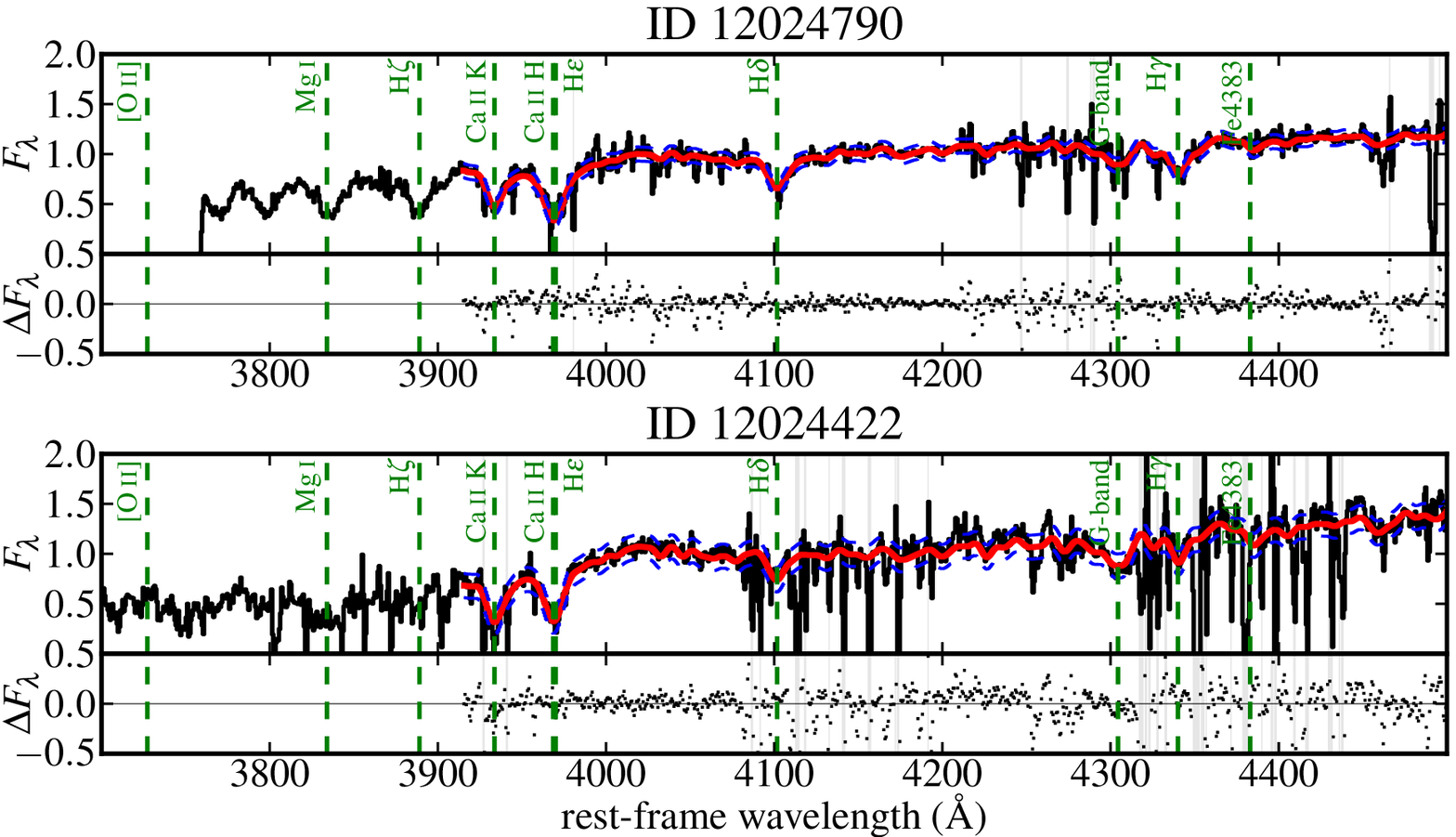}
  \caption{\textsc{pPXF} fitting results for the targets in the slit 2 of the observations. On top panels black thin solid lines show the spectra, red thick solid lines represent \textsc{pPXF} fits, and blue dashed lines are 1$\sigma$ deviations from the \textsc{pPXF} fits. On bottom panels we have plotted the residuals with black dots. Units are arbitrary. Grey regions indicate those which were masked in the \textsc{pPXF} fits. We have marked some spectral features with green dashed vertical lines.}
  \label{fig:spectra-slit2}
\end{figure*}

\begin{figure*}
  \includegraphics{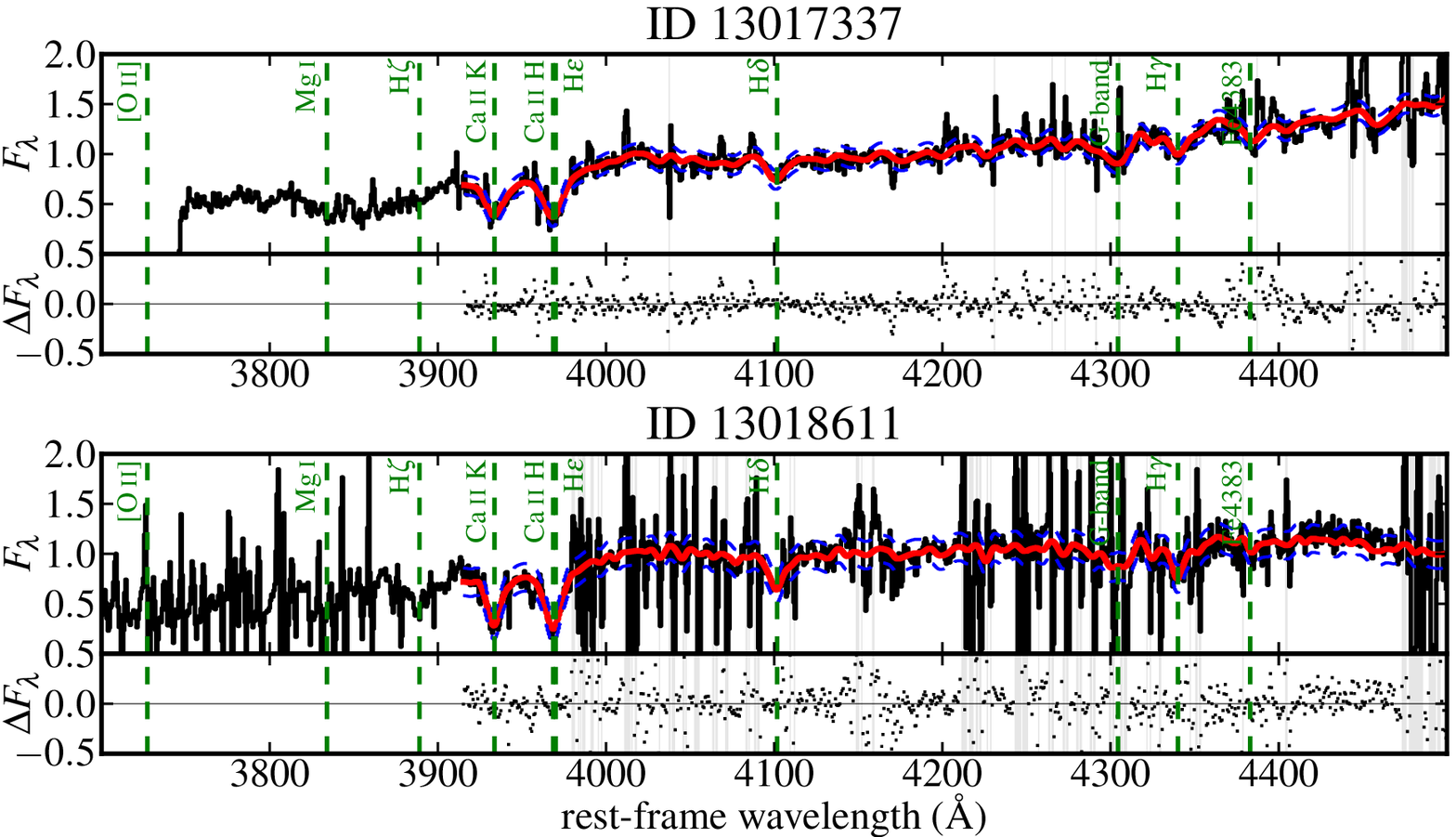}
  \caption{\textsc{pPXF} fitting results for the targets in the slit 3 of the observations. On top panels black thin solid lines show the spectra, red thick solid lines represent \textsc{pPXF} fits, and blue dashed lines are 1$\sigma$ deviations from the \textsc{pPXF} fits. On bottom panels we have plotted the residuals with black dots. Units are arbitrary. Grey regions indicate those which were masked in the \textsc{pPXF} fits. We have marked some spectral features with green dashed vertical lines.}
  \label{fig:spectra-slit3}
\end{figure*}

We show in Figs~\ref{fig:spectra-slit1}--\ref{fig:spectra-slit3} the spectra obtained after the data reduction explained in Section~\ref{subsec:reduction}, together with the spectral fit obtained from \textsc{pPXF} detailed in Section~\ref{subsec:velocity-dispersions}.

\newpage 

\section{Structural parameters for the \lowercase{\boldmath $z \sim 1$} sample} \label{ap:sample-z1}

Table~\ref{tab:sample-z1} contains a compilation of structural parameters for the sample of massive ETGs at $z \sim 1$ analysed in this paper.

\addtolength{\tabcolsep}{4.6mm} 
\begin{table*}
    \centering
    \begin{minipage}{147.5mm}
        \caption{Compilation of structural parameters for the $z \sim 1$ sample.
        (1) References: [0] this work; [1] \citet{2011ApJ...738L..22M}; [2] \citet{2014ApJ...783..117B}; [3] \citet{2010ApJ...717L.103N} \citep[extracted from the compilation made by][]{2013ApJ...771...85V}; [4] \citet{2008ApJ...688...48V} and \citet{2006ApJ...644...30B} \citep[extracted from the compilation made by][]{2013ApJ...771...85V}.
        (2) Galaxy identifications as given in the references.
        (3) Spectroscopic redshifts.
        (4) S\'ersic indices.
        (5) Circular effective (half-light) radii.
        (6) Stellar masses (adapted to Salpeter IMF).
        (7) Velocity dispersions within the effective (half-light) radius.}
        \label{tab:sample-z1}
        \begin{tabular}{@{}lcd{1.5}cccc@{}}
        \hline
        Reference & ID       & \multicolumn{1}{c}{$z$} & $n$ & \re    & \Mstar              & \sigmae\\
                  &          &                         &     & (\kpc) & ($10^{11} \ \Msun$) & (\kms)\\
        (1)       & (2)      & \multicolumn{1}{c}{(3)} & (4) & (5)    & (6)                 & (7)\\
        \hline
        0         & 13004276 & 0.97533                 & 5.1 & 1.46   & 2.0                 & $340 \pm 16$\\
        0         & 12028173 & 0.95764                 & 5.9 & 1.42   & 1.2                 & $228 \pm 17$\\
        0         & 12024790 & 0.96572                 & 4.4 & 0.54   & 0.9                 & $261 \pm 15$\\
        0         & 12024422 & 1.02874                 & 4.8 & 1.57   & 1.2                 & $239 \pm 18$\\
        0         & 13017337 & 0.97242                 & 4.3 & 1.37   & 1.8                 & $291 \pm 25$\\
        0         & 13018611 & 1.07939                 & 4.7 & 1.19   & 3.0                 & $161 \pm 16$\\
        1         & 12024321 & 0.91592                 & 6.5 & 2.46   & 0.9                 & $162 \pm 12$\\
        1         & 12019899 & 0.93259                 & 6.3 & 0.47   & 1.1                 & $245 \pm 18$\\
        1         & 12024453 & 0.90564                 & 4.8 & 1.60   & 1.7                 & $184 \pm 10$\\
        2         & 51106    & 1.013                   & 5.3 & 5.99   & 3.3                 & $252 \pm 37$\\
        2         & 28739    & 1.029                   & 3.5 & 1.98   & 1.6                 & $238 \pm 11$\\
        2         & 54891    & 1.081                   & 3.2 & 1.27   & 1.0                 & $232 \pm 37$\\
        2         & 31377    & 1.085                   & 4.9 & 4.88   & 1.2                 & $133 \pm 18$\\
        2         & 13393    & 1.097                   & 3.5 & 7.18   & 2.6                 & $175 \pm 21$\\
        2         & 16343    & 1.098                   & 8.0 & 1.95   & 2.0                 & $290 \pm 8 \ \ $\\
        3         & E1       & 1.054                   & 4.0 & 6.44   & 2.1                 & $204 \pm 22$\\
        4         & 761      & 1.01                    & 4.0 & 3.77   & 6.2                 & $377 \pm 40$\\
        4         & 1559     & 0.94                    & 4.0 & 1.66   & 1.1                 & $179 \pm 13$\\
        4         & 1706     & 0.91                    & 4.0 & 2.23   & 2.7                 & $217 \pm 13$\\
        4         & 1        & 1.09                    & 4.0 & 2.83   & 2.7                 & $233 \pm 17$\\
        4         & 2        & 0.96                    & 4.0 & 2.30   & 3.5                 & $202 \pm 10$\\
        4         & 3        & 1.04                    & 4.0 & 1.00   & 1.2                 & $302 \pm 34$\\
        4         & 4        & 0.96                    & 4.0 & 6.84   & 7.8                 & $337 \pm 19$\\
        4         & 13       & 0.98                    & 4.0 & 2.20   & 3.2                 & $249 \pm 11$\\
        4         & 14       & 0.98                    & 4.0 & 2.80   & 1.4                 & $199 \pm 24$\\
        4         & 18       & 1.10                    & 4.0 & 3.97   & 5.7                 & $327 \pm 37$\\
        4         & 20       & 1.02                    & 4.0 & 2.24   & 2.6                 & $201 \pm 17$\\
        \hline
        \end{tabular}
        \\ \\ \\ \bsp 
    \end{minipage}
    \label{lastpage} 
\end{table*}
\addtolength{\tabcolsep}{-4.6mm} 



\end{document}